\documentclass[aps,prl,twocolumn,groupedaddress,nofootinbib,longbibliography]{revtex4-1}

\usepackage{graphicx}% Include figure files
\usepackage{dcolumn}% Align table columns on decimal point
\usepackage{multirow}
\usepackage{siunitx}
\usepackage{amsmath}
\usepackage[hidelinks]{hyperref}

\newcommand{\angstrom}{\mbox{\normalfont\AA}}
\newcommand{\xLi}{x_\mathrm{Li}}
\newcommand{\lipath}[4]{$^\mathrm{#1}\mathrm{#2}_{#3}^\mathrm{#4}$}
\newcommand{\lmo}{Li$_y$Mn$_2$O$_4$}
\newcommand{\lto}{Li$_{4+3z}$Ti$_5$O$_{12}$}

\usepackage[normalem]{ulem}
\usepackage{color}

\newcommand{\done}[1]{\textcolor[rgb]{1.0,0.4,0.0}{$=$ #1}}
\newcommand{\noted}[1]{\textcolor[rgb]{0.0,0.8,0.8}{\textsc{[#1]}}}

\newcommand{\chem}[1]{\ensuremath{\mathrm{#1}}}

\makeatletter\newcommand{\etal}{\emph{et al}\@ifnextchar.{}{.\ }}\makeatother

\begin{document}

\title{Interfacial Strain Effects on Lithium Diffusion Pathways in the Spinel Solid Electrolyte Li-Doped \texorpdfstring{MgAl$_2$O$_4$}{MgAl2O4}}

\author{Conn O'Rourke}
  \email{c.o'rourke@bath.ac.uk}
  \affiliation{Department of Chemistry, University of Bath, Bath, BA2 7AX}

\author{Benjamin J. Morgan}
  \email{b.j.morgan@bath.ac.uk}
  \affiliation{Department of Chemistry, University of Bath, Bath, BA2 7AX}

% uncomment these lines to hide comments in the manuscript
\renewcommand{\done}[1]{}
\renewcommand{\noted}[1]{}

\date{\today}

\begin{abstract}
(Li,Al)--co-doped magnesium spinel (Li$_x$Mg$_{1-2x}$Al$_{2+x}$O$_4$) is a  solid lithium-ion electrolyte with potential use in all-solid-state lithium-ion batteries. 
The spinel structure means that interfaces with spinel electrodes, such as \lmo{} and \lto{}, may be \emph{lattice-matched}, with potentially low interfacial resistances. Small lattice parameter differences across a lattice-matched interface are unavoidable, causing residual epitaxial strain. This strain potentially modifies lithium diffusion near the electrolyte--electrode interface, contributing to interfacial resistance. 
Here we report a density functional theory study of strain effects on lithium diffusion pathways for (Li,Al)--co-doped magnesium spinel, for $\xLi=0.25$ and $\xLi=0.5$.
We have calculated diffusion profiles for the unstrained materials, and for isotropic and biaxial tensile strains of up to $6\%$, corresponding to $\left\{100\right\}$ epitaxial interfaces with Li$_y$Mn$_2$O$_4$ and Li$_{4+3z}$Ti$_5$O$_{12}$.
We find that isotropic tensile strain reduces lithium diffusion barriers by as much as $0.32\,\mathrm{eV}$, with typical barriers reduced by $\sim0.1\,\mathrm{eV}$. This effect is associated with increased volumes of transitional octahedral sites, and broadly follows qualitative changes in local electrostatic potentials. 
For biaxial (epitaxial) strain, which more closely approximates strain at a lattice-matched electrolyte--electrode interface, changes in octahedral site volumes and in lithium diffusion barriers are much smaller than under isotropic strain. Typical barriers are reduced by only $\sim0.05\,\mathrm{eV}$. Individual effects, however, depend on the pathway considered and the relative strain orientation.
These results predict that isotropic strain strongly affects ionic conductivities in (Li,Al)--co-doped magnesium spinel electrolytes, and that tensile strain is a potential route to enhanced lithium transport. For a \emph{lattice-matched} interface with candidate spinel-structured electrodes, however, epitaxial strain has a small, but complex, effect on lithium diffusion barriers.
\end{abstract}

\maketitle

\section{\label{sec:introduction}Introduction}
Lithium-ion batteries are widely used for energy storage. Historically, the principal applications have been in consumer electronics, but lithium-ion batteries are increasingly finding use in other sectors, such as electric vehicles and grid-scale storage. This success has been tempered by concerns about the stability and safety of the liquid electrolytes found in commercial lithium-ion batteries. Organic liquid electrolytes are flammable, and present an explosion risk if a cell were to short-circuit or suffer mechanical failure. One proposed solution is to replace the liquid electrolytes with electrochemically inert solid lithium-ion conductors. This would eliminate the explosion risk, remove the need for costly protection circuitry, and allow possible battery miniaturization \cite{Knauth_SolStatIonics2009,BachmanEtAl_ChemRev2016, ManthiramEtAl_NatRevMater2017}. 

A number of promising lithium-ion solid electrolytes exist; some with ionic conductivities comparable to those of liquid electrolytes \cite{KamayaEtAl_NatMater2011, BachmanEtAl_ChemRev2016, ManthiramEtAl_NatRevMater2017}. 
For a solid lithium-ion electrolyte to be commercially viable it must be electrically and chemically stable under cell-operating conditions, and have  high ionic and negligible electronic conductivities. 
In an all--solid-state lithium-ion battery, device performance also depends on the  interfacial resistance between the electrolyte and electrodes \cite{SanthanagopalanEtAl_JPhysChemLett2014}. 
If a solid electrolyte--electrode pair have crystal structures or lattices that are incommensurate, the electrolyte--electrode interface is expected to be \emph{mismatched}. Lithium-diffusion pathways across these mismatched interfaces may be highly tortuous, resulting in large interfacial resistances \cite{RoscianoEtAl_PhysChemChemPhys2013}. 

One strategy for developing all--solid-state batteries with low interfacial resistances is to choose electrolyte--electrode combinations that are \emph{lattice-matched} \cite{Thackeray_Patent1985, RoscianoEtAl_PhysChemChemPhys2013}.  
If the cathode, electrode, and anode have mutually compatible crystal structures that allow epitaxially coherent interfaces, the diffusion pathways across these solid--solid interfaces are expected to be relatively unobstructed, giving low interfacial resistances.
Lattice matching between electrodes and a solid lithium-ion electrolyte was first described by Thackeray and Goodenough in 1985 \cite{Thackeray_Patent1985}, who suggested using spinel-structured materials for each of the anode, cathode, and electrolyte. 
Spinel structured electrodes, such as \lmo{} lithium manganate cathodes and \lto{} lithium titanate anodes are well known, and are already used in commercial batteries. Spinel-structured electrolytes, however, have long proved elusive. In 2013, Rosciano \etal synthesised a new class of spinel-structured lithium-ion electrolytes based on (Li,Al)--co-doped magnesium spinel (Li$_x$Mg$_{1-2x}$Al$_{2+x}$O$_4$) \cite{RoscianoEtAl_PhysChemChemPhys2013, BlaakmeerEtAl_JPhysChemC2015, DjenadicEtAl_SolStatIonics2016}. This development has revived interest in the possibility of all-solid-state lithium-ion batteries constructed to have a coherent face-centred--cubic (fcc) oxide lattice shared across the anode, electrolyte, and cathode. 

Lattice matched electrode--electrolyte interfaces are also interesting in the context of protective coatings for cathodes used with conventional electrolyte chemistries. Coating cathode materials with solid electrolyte thin films can enhance cell performance and increase operating lifetimes, by inhibiting electrolyte--electrode side-reactions, and improving the mechanical resilience of the cathode during lithium insertion and extraction \cite{YiEtAl_Ionics2009,ChenEtAl_JMaterChem2010,AykolEtAl_NatureComm2016,ZuoEtAl_JAllCom2017}. Solid thin film coatings have also been used in solid-state lithium-ion batteries as a separating layer between the cathode and a solid electrolyte, to reduce interfacial resistance \cite{OhtaEtAl_AdvMater2006,OhtaEtAl_ElectrochemComm2007,TakadaAndOhno_FrontEnergyRes2016}. In both these cases, the interfacial resistance between the cathode and the thin-film coating can limit cell performance. Using lattice-matched solid electrolytes as thin film surface layers is one strategy to design low resistance cathode--thin-film interfaces. Epitaxial matching between a spinel-structured cathode and a spinel-structured surface coating has been demonstrated by Li \etal who have coated spinel \chem{LiMn_2O_4} with \chem{Li_4Ti_5O_{12}}, giving a coherent epitaxial interface between the two materials \cite{LiEtAl_ACSApplMaterInt2014}.

Perfect lattice-matching between structurally compatible electrode--electrolyte pairs is not achievable in practice, and any coherent lattice-matched interface will exhibit some residual strain. 
For example, the lattice parameter of \chem{MgAl_2O_4} differs from those of \lmo{} and \lto{} by $0.04\%$--$2.3\%$ and $4.0\%$, respectively \cite{BergAndThomas_SolStatIonics1999, WagemakerEtAl_AdvMater2006}. (Li,Al)-codoping of \chem{MgAl_2O_4} causes the lattice parameter to contract as the lithium concentration is increased (see Table \ref{tab:lattice_parameters}). The degree of interfacial strain therefore depends not only on the choice of interfacing electrode, but also on the electrolyte stoichiometry. 

Residual strain at electrolyte--electrode interfaces may not be without consequences. Lattice strains modify local atomic geometries and distort the potential energy surface that defines lithium diffusion pathways. This affects diffusion barrier heights, and hence ionic conductivities. The effect of interfacial strain on ionic conductivities has been widely studied in fuel cell materials \cite{SchichtelEtAl_PhysChemChemPhys2009, Rupp_SolStatIonics2012, WenEtAl_JMaterChemA2015, Yildiz_MRSBull2014, AydinEtAl_PhysChemChemPhys2013, ShenEtAl_RSCAdv2014, FluriEtAl_NatureComm2016, FerraraEtAl_PhysChemChemPhys2016}, motivated by the possibility of straining thin film electrolytes to enhance their oxide-ion conductivities, and hence reduce device operating temperatures. More recently, the concept of ``strain engineering'' has also been considered as a strategy for enhancing lithium conductivity in lithium-ion electrodes and electrolytes \cite{YanEtAl_FunctMaterLett2012,LeeEtAl_ApplPhysLett2012,NingEtAl_SolStatIonics2014, WeiEtAl_CrystGrowthDes2015, Tealdi_JMaterChemA2016, MoradabadiAndKaghazchi_PhysRevAppl2017, JiaEtAl_RSCAdv2017,MuralidharanEtAl_ACSNano2017,MoradabadiEtAl_arXiv2017,SagotraAndCazorla_ACSApplMaterInt2017}. The effect of interfacial strain is particularly pertinent for an electrolyte such as (Li,Al)--co-doped \chem{MgAl_2O_4}, where interest is motivated by the possibility of lattice-matching with spinel-structured electrolytes. 

\begin{table}[t]
  \begin{center}
    \begin{tabular}{llcll}
        System                    & $a_\mathrm{expt.}$ [\angstrom] & Ref. & \multicolumn{2}{l}{$a_\mathrm{PBEsol}$ [\angstrom]} \\ \hline
        Li$_{0.28}$Mn$_2$O$_4$        & 8.043 & \cite{BergAndThomas_SolStatIonics1999} & \\
        Li$_{0.74}$Mn$_2$O$_4$        & 8.144 & \cite{BergAndThomas_SolStatIonics1999} &  \\
        LiMn$_2$O$_4$ & 8.221 & \cite{BergAndThomas_SolStatIonics1999} & \\ \hline
        \lto{}        & 8.360 & \cite{WagemakerEtAl_AdvMater2006} &   \\ \hline
        \chem{MgAl_2O_4} ($\xLi=0$) & 8.04 & \cite{RoscianoEtAl_PhysChemChemPhys2013} &  8.051\\
        \chem{Li_{0.25}Mg_{0.5}Al_{2.25}O_4} ($\xLi=0.25$) & 7.96 & \cite{RoscianoEtAl_PhysChemChemPhys2013} & 7.996 \\
        \chem{Li_{0.5}Al_{2.5}O_4}  ($\xLi=0.5$)& 7.89 & \cite{RoscianoEtAl_PhysChemChemPhys2013} & 7.907 \\ 
    \end{tabular}
  \caption{\label{tab:lattice_parameters}Lattice parameters for lithium manganate and lithium titanate spinels, and for Li$_x$Mg$_{1-2x}$Al$_{2+x}$O$_4$, from experiment and as calculated (PBEsol) for this work.}
  \end{center}
\end{table}

Computational modelling provides a powerful tool for studying how factors such as stoichiometry or strain can affect lithium diffusion. To date, the only computational study of lithium transport in (Li,Al)--co-doped magnesium spinel was performed by Mees \etal \cite{MeesEtAl_PhysChemChemPhys2014}. These authors modelled lithium diffusion at dopant concentrations of $\xLi=0.125$, $\xLi=0.25$, and $\xLi=0.50$, by calculating lithium-ion diffusion barriers using density-functional theory (DFT) nudged--elastic-band (NEB) calculations, These diffusion barriers were then used in kinetic Monte Carlo (kMC) simulations to model long time-scale diffusion. 
These kMC simulations predicted that increasing $\xLi$ from $0.25$ to $0.50$ causes the lithium diffusion coefficient to increase by $\times 10^2$, with a corresponding decrease in activation energy from $0.61\,\mathrm{eV}$ to $0.45\,\mathrm{eV}$.\footnote{Interestingly, the local DFT-NEB diffusion barriers for lithium motion calculated by Mees \etal \cite{MeesEtAl_PhysChemChemPhys2014} are broadly independent of dopant concentration, suggesting that the dominant effect of stoichiometry on the lithium diffusion coefficient is due to changes in the connectivity of available diffusion pathways.} 

Despite this predicted increase in lithium diffusion coefficient with (Li,Al)-dopant concentration, Mees \etal suggested that the optimal composition for use with the electrodes \lmo{} or \lto{} is $\xLi\approx0.3$. This effective limit was proposed with two considerations in mind. First, that high lithium content may promote phase separation to poorly conducting \chem{LiAl_5O_8} and \chem{MgAl_2O_4}. Second, because the bulk lattice parameter of (Li,Al)--co-doped spinel decreases as $\xLi$ increases, a higher lithium content corresponds to a \emph{larger} lattice-parameter mismatch with the spinel electrolytes \lmo{} and \lto{}. This, in turn, was suggested to be likely to increase interfacial resistance, and reduce overall device performance. Subsequent experimental work has shown that (Li,Al)--co-doped spinel can be synthesised at high dopant concentrations (up to $\xLi=0.4$) without phase separation \cite{DjenadicEtAl_SolStatIonics2016}. The first concern regarding phase stability then, may be (at least partially) avoided through careful synthesis. The second question, however; the effect of epitaxial strain on lithium diffusion, in particular as $\xLi$ increases; remains open. 

Here we report density functional theory (DFT) climbing-image nudged--elastic-band (CI-NEB) \cite{HenkelmanEtAl_JChemPhys2000} calculations of lithium diffusion pathways in unstrained, isotropically strained, and epitaxially (biaxially) strained (Li,Al)--co-doped MgAl$_2$O$_4$. We have performed these calculations to better understand the consequences of strain at a hypothetical lattice-matched interface between (Li,Al)--co-doped spinel and the electrodes \lmo{} and \lto{}. 
Our calculations assume an implicit coherent interface across the full range of dopant stoichiometries considered. Increasing $\xLi$ increases the degree of lattice mismatch with respect to both candidate electrodes. At high mismatch values it becomes increasingly likely that dislocations or other extended defects form, which relieve the interfacial strain, and which may have additional effects on lithium transport. In practice, high quality interfaces may require thin films of electrolyte, which are more able to preserve epitaxy under large lattice misfits \cite{AllimiEtAl_ApplPhysLett2008}.

We find that in (Li,Al)--co-doped spinel, isotropic tensile strain reduces lithium diffusion barriers, from $\sim0.4\,\mathrm{eV}$ to $\sim0.25\,\mathrm{eV}$ under strains of up to $6\%$. This effect is correlated with changes in the volume of the octahedral site at the midpoint of each diffusion pathway, and differences in diffusion barriers are approximately correlated with differences in electrostatic potential along specific paths. For anisotropic strain, which approximates an ideal coherent electrolyte--electrode interface, the transition site volume changes are much smaller, and the corresponding effect on the lithium-diffusion barrier height is weaker. While the effect of epitaxial strain on most barriers is a small reduction ($\sim0.05\,\mathrm{eV}$ at 6\% strain) the quantitative details are more complex than in the isotropic strain case: changed to barrier heights depend on the choice of diffusion path being considered, and the orientation relative to the applied strain. These results indicate that isotropic strain is expected to have a large impact on ionic conductivities in (Li,Al)--co-doped spinel electrolytes, with tensile strain being a potential route to enhanced lithium transport. For ``lattice-matched'' electrolyte--electrode systems, however, providing interfaces remain coherent, the residual epitaxial strain is not expected to strongly affect lithium diffusion barriers.

\section{\label{Method}Methods}

\begin{figure}[tb]
  \centering
  \resizebox{7cm}{!}{\includegraphics*{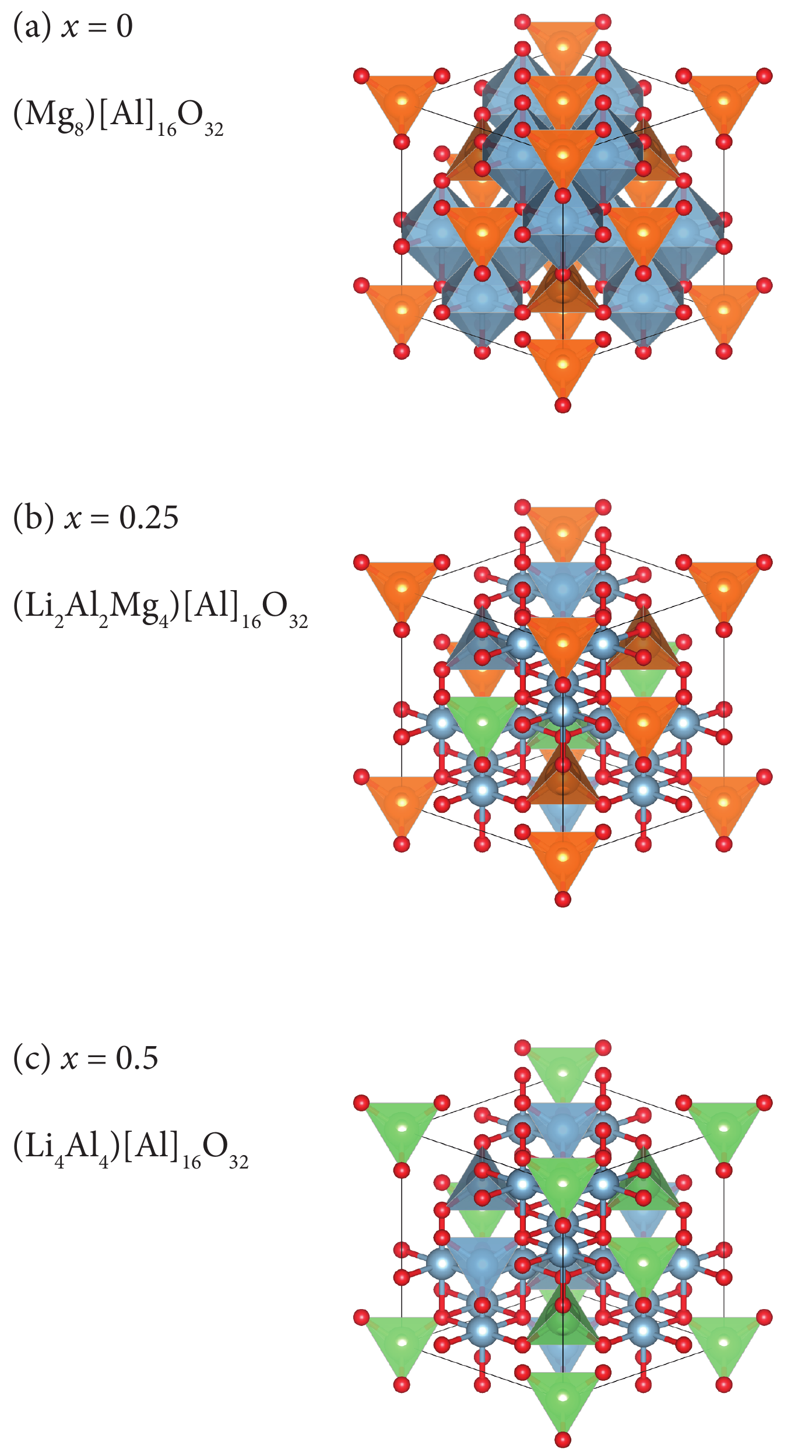}} %
    \caption{\label{fig:doped_spinel_structures} Example 56 atom unit cells for undoped and (Li,Al)-doped spinel structures with composition (Al$_{x}$Mg$_{1-2x}$Li$_{x}$)[Al$_{2}$]O$_{4}$. (a) The undoped parent spinel ($\xLi=0$): \chem{(Mg_8)[Al_{16}]O_{32}}, (b) 50\% doped ($\xLi=0.25$): \chem{(Al_{2}Li_{2}Mg_{4})[Al_{16}]O_{32}}, (c) Fully doped ($\xLi=0.5)$: \chem{(Al_{4}Li_{4})[Al_{16}]O_{32}}. Mg atoms are orange, Al atoms are blue, Li atoms are green, and O atoms are red.}
\end{figure}

All calculations were performed using the plane-wave DFT code \texttt{VASP} \cite{KresseAndFurthmuller_PhysRevB1996, KresseAndFurthmuller_CompMaterSci1996}. Exchange and correlation effects were approximated by the revised Perdew-Burke-Ernzerhof generalized gradient approximation (GGA) PBEsol functional \cite{PerdewEtAl_PhysRevLett2008}. The pseudopotential method was used in the form of projector augmented wave (PAW) pseudopotentials to treat core electrons \cite{Blochl_PhysRevB1994} with 3 valence electrons for Al (3\emph{s}$^2$3\emph{p}$^1$), 1 valence electron for Li (2\emph{s}$^1$), 6 valence electrons for O (2\emph{s}$^2$2\emph{p}$^6$), and 2 valence electrons for Mg (3\emph{s}$^2$).
All calculations were performed on 56 atom cells, and $k$-space sampling used a ($3\times3\times3$) Monkhorst-Pack grid. The plane-wave cut-off energy was \SI{600}{\electronvolt} for calculations with variable cell shape and volume, and \SI{550}{\electronvolt} for calculations with fixed cell shape. Geometry optimisations, including the CI-NEB calculations were, were deemed converged when the forces on ions were less than \SI{1e-2}{\electronvolt\per\angstrom}.

The conventional spinel structure belongs to the Fd$\bar{3}$m spacegroup, and has 56 ions in the unit cell, with stoichiometry 
(A$^{2+}$)[B$_2^{3+}$]O$_4$. In the normal magnesium-spinel structure, \chem{(Mg)[Al_{2}]O_{4}}, \chem{Mg^{2+}} cations occupy the tetrahedral 8a Wycoff positions, \chem{Al^{3+}} cations occupy the octahedral 16d positions, and \chem{O^{2-}} anions occupy the 32e positions \cite{SickafusEtAl_JAmCermSoc1999, HillEtAl_PhysChemMiner1979}. Upon co-doping with \{Li,Al\}, pairs of \chem{Mg^{2+}} cations are substituted in equal proportion by Li$^+$ and Al$^{3+}$, to maintain charge neutrality, giving a composition of (Al$_{x}$Mg$_{1-2x}$Li$_{x}$)[Al$_{2}$]O$_{4}$ \cite{RoscianoEtAl_PhysChemChemPhys2013, MeesEtAl_PhysChemChemPhys2014}. The maximum possible dopant content is $\xLi=0.5$ (Fig.\ \ref{fig:doped_spinel_structures}).

\begin{figure}[tb]
  \centering
  \resizebox{5cm}{!}{\includegraphics*{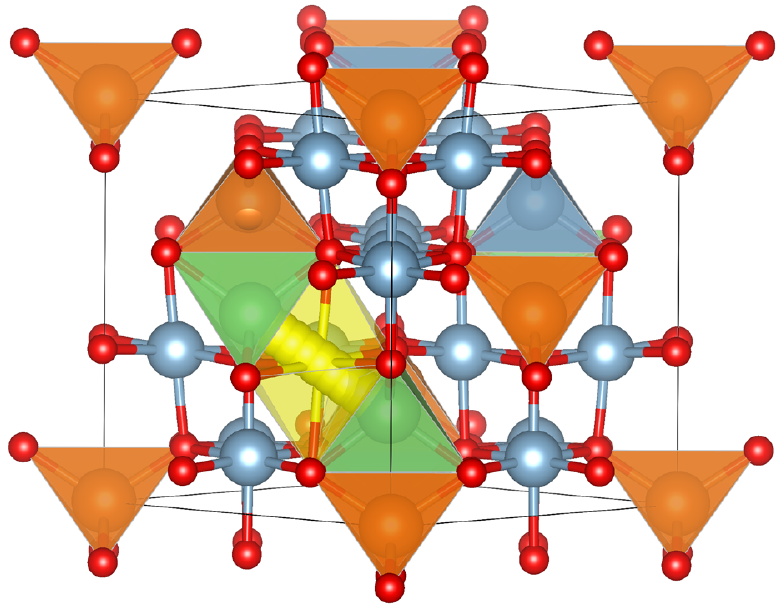}} %
    \caption{\label{fig:POSCAR_pathway}Li$^+$ diffusion pathway between neighbouring A cation sites via the 16c octahedral vacancy (yellow).}
\end{figure}

For our calculations, we consider dopant concentrations of $\xLi=0.25$ and $\xLi=0.5$. Substitutional doping at the Mg sites lowers the crystal symmetry, making the A-site cations non-equivalent. To account for this A-site disorder we identified all symmetry inequivalent structures at each dopant concentration using the \texttt{bsym} code \cite{Morgan_JOpenSourceSoftware_bsym}. In a 56 atom cell, this gives 7 structures at $\xLi=0.25$ and 4 structures at $\xLi=0.5$. Fig.\ \ref{fig:doped_spinel_structures} shows examples of these doped structures alongside the undoped magnesium-spinel structure. Li$^+$ diffusion between the tetrahedral A-sites proceeds via 8a--16c--8a paths, with the vacant 16c octahedra connecting the end-point tetrahedra (Fig.\ \ref{fig:POSCAR_pathway}). 

In this study we focus on vacancy--mediated lithium-ion transport, following the approach of Mees \etal \cite{MeesEtAl_PhysChemChemPhys2014}. For each inequivalent doped structure, we generated the full set of possible lithium-diffusion pathways by removing, in turn, each of the A-site lithium ions. At a dopant content of $\xLi=0.25$, three of the seven initial symmetry inequivalent structures have viable lithium diffusion pathways, where a lithium ion can move from an occupied A site, through an octahedral 16c site, into a previously unoccupied destination A-site. These three structures have one unique pathway each, which can be traversed in either direction. At a dopant content of $\xLi=0.5$, three of the four symmetry inequivalent structures have viable lithium-diffusion pathways. Two of these structures have a single pathway, and the third structure has two. 

To sample these different diffusion pathways, we consider all three inequivalent paths at $\xLi=0.25$, and three of the available paths at $\xLi=0.5$. For each lithium concentration, this gives a set of pathways that includes all possible combinations of initial and final A-site coordination environments within a 56 atom unit cell. For each candidate pathway, we have performed a series of CI-NEB calculation to evaluate the potential energy profile of a diffusing lithium ion \cite{HenkelmanEtAl_JChemPhys2000}.
We denote specific pathways using $S_x^p$, where $S$ describes the target strain (U = unstrained, LM = lithium manganate lattice parameters, LT = lithium titanate lattice parameters), $x$ is the dopant concentration $\xLi$, and $p$ enumerates the pathways. An additional superscript R indicates a ``reversed'' pathway in cases where the diffusion profile is not symmetric. To model isotropic strain, all three lattice parameters are equally scaled to match that of the target electrode. For anisotropic epitaxial strain, only two of the lattice parameters are adjusted to match the target lattice. The perpendicular lattice parameter is adjusted to minimise the total cell energy.

\begin{figure}[htb]
  \centering
  \resizebox{8.5cm}{!}{\includegraphics*{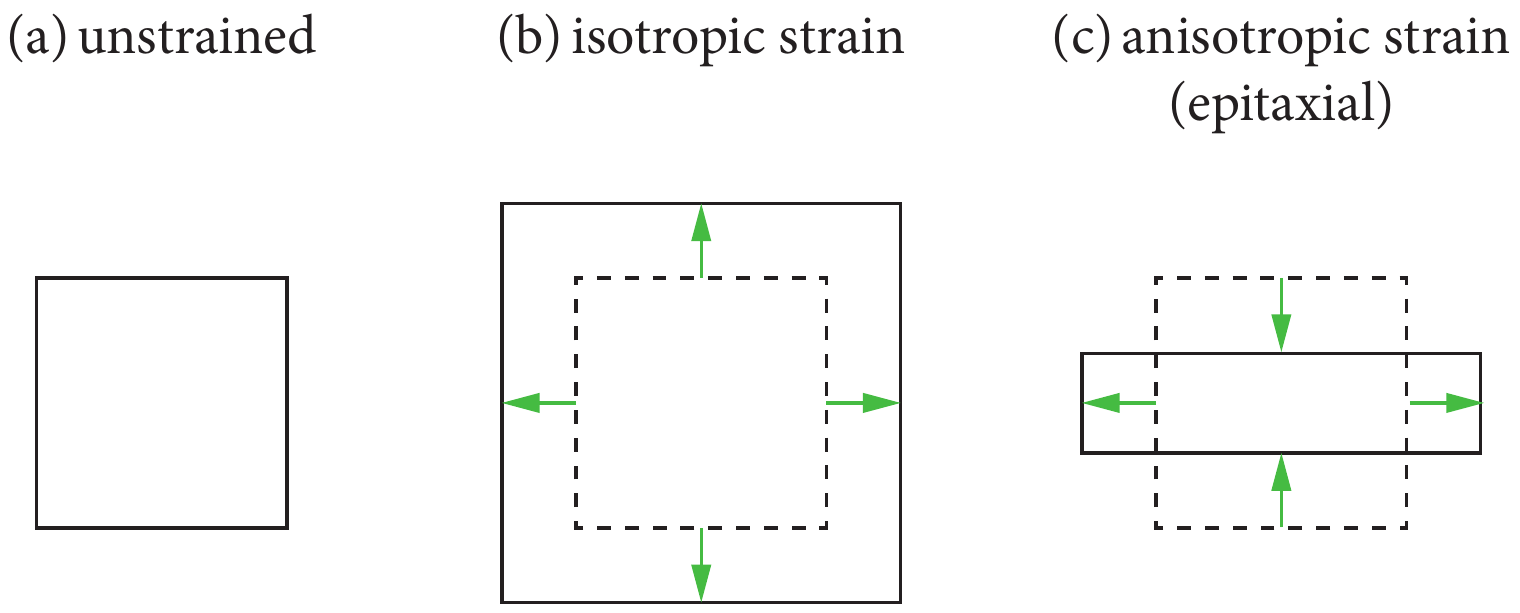}} %
    \caption{\label{fig:strain}2D schematic illustrating the unstrained, isotropically strained, and anisotropically strained calculations. (a) Unstrained: all lattice parameters are relaxed during the DFT calculation ($a,b,c = a^\mathrm{DFT}$) (b) Isotropic strain: all three lattice parameters are scaled to match those of the target electrode ($V_\mathrm{Cell} = V_\mathrm{electrode}$) (c) Anisotropic strain: the two in-plane lattice parameters are scaled to match those the target electrode, and the perpendicular lattice parameter is relaxed ([$a,b = a^\mathrm{electrode}$, $c=c^\mathrm{DFT}$], [$b,c = a^\mathrm{electrode}$, $a=a^\mathrm{DFT}$], [$a,c = a^\mathrm{electrode}$, $b=b^\mathrm{DFT}$]) }
\end{figure}

The different strain protocols; unstrained, isotropic, and epitaxial; are illustrated schematically in Fig.~\ref{fig:strain}. In all cases, we consider strains corresponding to model $\{100\}_\mathrm{A}|\{100\}_\mathrm{B}$ coherent interfaces between the doped (Mg,Al)-spinel and the target electrode (lithium manganate or lithium titanate spinel). \lto{} is a ``zero-strain'' material, whose lattice parameters change by $<0.1\%$ upon lithium intercalation and extraction \cite{RonciEtAl_JPhysChemB2002,WagemakerEtAl_AdvMater2006, MorganEtAl_JMaterChemA2016}. The lattice parameters of \lmo{}, however, expand and contract on lithium insertion and extraction, varying from $8.04\,\mathrm{\angstrom}$ ($0.04\%$ strain versus \chem{MgAl_2O_4}) at $\xLi=0.28$, to $8.36\,\mathrm{\angstrom}$ ($2.3\%$ strain versus \chem{MgAl_2O_4}) at $\xLi=1.0$ \cite{BergAndThomas_SolStatIonics1999}. For our calculations we consider an intermediate \lmo{} lattice parameter of $8.144\,\mathrm{\angstrom}$ ($1.3\%$ strain versus \chem{MgAl_2O_4}), which corresponds to a stoichiometry of $\xLi\approx0.74$. This gives three strain values that span our range of interest, for each diffusion pathway. The behaviour at intermediate strains, such as those corresponding to fully inserted or delithiated \lmo{}, can be extrapolated by interpolating the calculated data. 

In addition to potential energy profiles from our CI-NEB calculations, we also present electrostatic potential profiles along each path. These are defined as the average electrostatic potential at the site of the mobile ion, evaluated at the optimised geometry of each NEB image. If a diffusing lithium ion is approximated as a $+1$ point charge, then the potential energy surface is characterised entirely by the electrostatic interactions between the mobile lithium ion and the surrounding lattice. Previous computational studies of lithium-ion diffusion in electrodes have shown that electrostatic potential profiles often give a good approximate description of diffusion barrier profiles \cite{MorganAndWatson_PhysRevB2010,TompsettEtAl_ChemMater2013,RongEtAl_JChemPhys2016}. In real systems, however, the interactions between mobile lithium ions and the host lattice are not purely electrostatic. Lithium ions also experience short-ranged repulsion from nearby lattice ions, due to overlapping valence electron densities. The detailed shape of a diffusion potential energy profile therefore depends on the balance of the electrostatic and short-ranged repulsive interactions.\footnote{For this analysis, we neglect lithium ion polarisation and dispersion interactions. Both contributions are expected to be small, due to the compact $1s^2$ electron configuration of Li$^+$.} In general, short-ranged repulsion is expected to be reduced under expansive strain. Our inclusion of electrostatic potential energy profiles is motivated, in part, by an interest in the extent to which this conceptually simple metric describes lithium-ion diffusion in (Li,Al)--co-doped spinel electrolytes \cite{TompsettEtAl_ChemMater2013, RongEtAl_JChemPhys2016}. 

To better understand the role of A-site cation disorder on lithium diffusion profiles, we also parametrise how the A-site coordination environments differ from those in undoped \chem{MgAl_2O_4}. We have followed the analysis of Mees \emph{et al.}, who noted that the energy of different co-doped spinel structures is correlated with the ``average'' local oxidation state for the A-site cations, $\mu_i$ \cite{MeesEtAl_PhysChemChemPhys2014}. For a specific A-site, $\mu_i$ is defined as the average formal oxidation state; $\mathrm{O}_\mathrm{Li}=+1, \mathrm{O}_\mathrm{Mg}=+2, \mathrm{O}_\mathrm{Al}=+3, \mathrm{O}_\mathrm{vacancy}=0$; of the ions occupying the central A-site, and the four nearest-neighbour A-sites:
\begin{equation}
\mu_i=\frac{1}{5}\left(\mathrm{O}_i+\sum_\mathrm{j\in\left\{nn_i^\mathrm{A}\right\}}\mathrm{O}_j\right)
\end{equation}
Structures with lower energies tend to have average $\mu$ values closer to $+2$, i.e.\ the value obtained for A-site Mg ions in the undoped parent structure. Deviations from this ``optimal'' average nearest-neighbour oxidation state can be quantified by calculating the root-mean-square difference between $\mu_i$ and the optimal value of $+2$ for all occupied A-sites in each structure:
\begin{equation}
\sigma_\mathrm{A}=\sqrt{\frac{\sum_i(\mu_{i}-2)^2}{7}}.
\label{eqn:sigma_A}
\end{equation}

Inputs and outputs for our \texttt{VASP} calculations are openly available under the CC-BY-SA-4.0 license \cite{ORourkeAndMorgan_SpinelNEBDataset2017}. This dataset includes Python scripts for extracting the NEB profiles, electrostatic potential profiles, and volumes of transitional octahedral sites from the DFT calculation outputs, and for calculating $\sigma_\mathrm{A}$ (Eqn.~\ref{eqn:sigma_A}). Analysis codes that produce Figs.\ \ref{fig:neb_pathways_iso}--\ref{fig:oct_vol_vs_delta_e_aniso} are available as a Jupyter notebook \cite{ORourkeAndMorgan_SpinelAnalysisNotebook2017,Hunter_matplotlib2007,KluyverEtAl_Jupyter2016}, published under the MIT license. Figs.\ \ref{fig:doped_spinel_structures} and \ref{fig:POSCAR_pathway} were generated using \texttt{VESTA} \cite{MommaEtAlJApplCryst2011}.

\section{\label{sec:results} {}Results}

%******************** Unstrained ************

\subsection{Unstrained (Li,Al)--co-doped spinel} 

\begin{figure*}[tb]
  \centering
  \resizebox{18cm}{!}{\includegraphics*{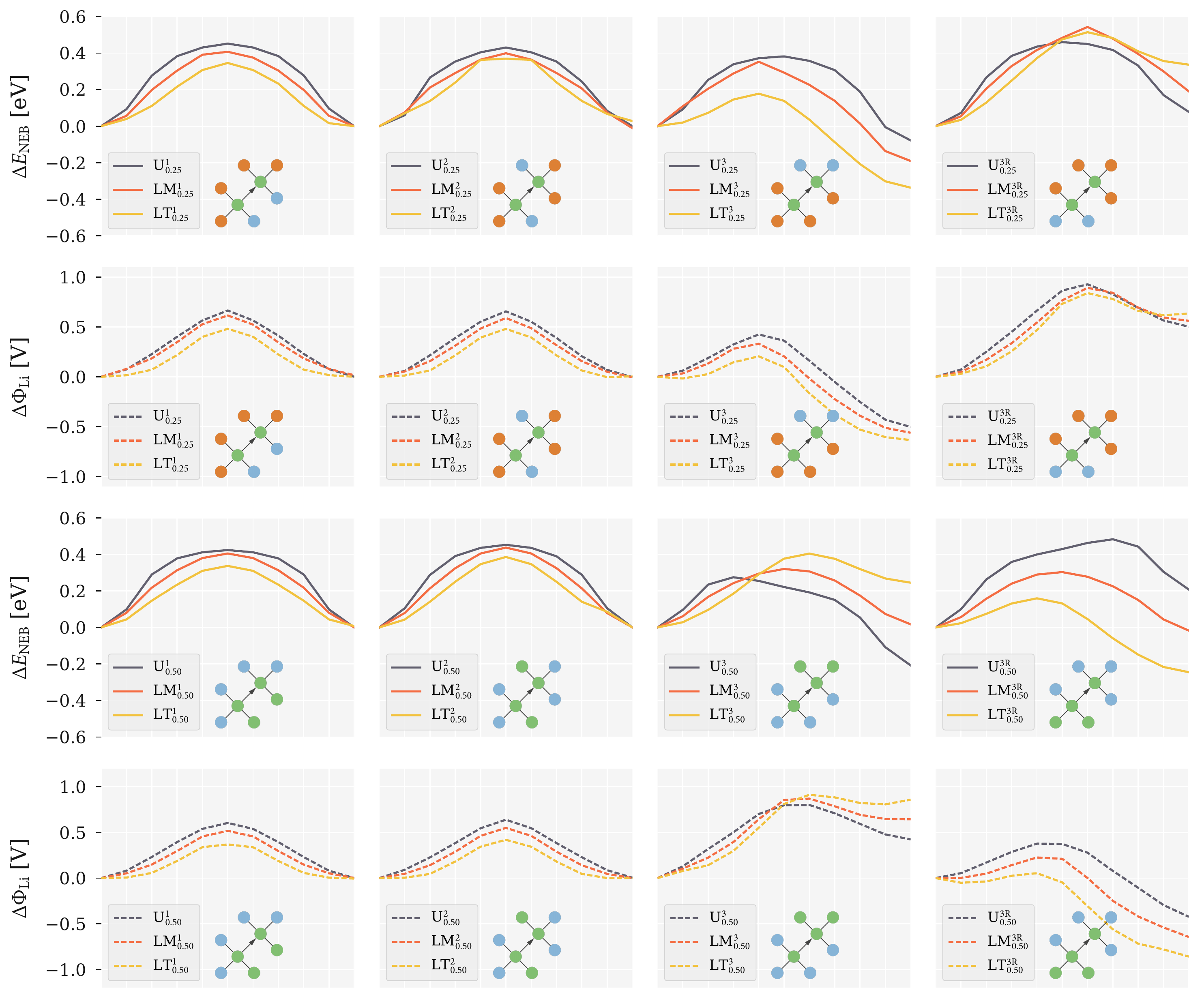}} %
    \caption{\label{fig:neb_pathways_iso}cNEB energy profiles and corresponding electrostatic potential profiles for unstrained and isotropically strained structures ($\xLi=0.25$ (top) and $\xLi=0.5$ (bottom)). Notation is as follows: $S_x^p$, where $S$ is strain (U = unstrained, LM = strained to lithium manganate lattice parameters, LT = strained to lithium titanate lattice parameters), $x$ is the dopant concentration and $p$ enumerates the pathways (where R refers to the same pathway in reverse). The inset schematic in each panel shows the coordination environments of the initial and final A-site positions. Li ions are green, Mg are orange and Al are blue. Electrostatic potential profiles (including the contribution from the mobile Li$^+$ ion) at the Li$^+$ site along the diffusion pathway are shown by the corresponding coloured dashed lines.}
\end{figure*}

We first consider lithium diffusion pathways in unstrained (Li,Al)--co-doped spinel, $(\mathrm{Al}_x\mathrm{Mg}_{1-2x}\mathrm{Li}_x)[\mathrm{Al}_2]\mathrm{O}_4$, using DFT-optimised lattice parameters for each dopant concentration.
The potential energy profiles for the minimum-energy pathways for lithium diffusion are shown in  Fig.~\ref{fig:neb_pathways_iso}, and the barrier heights are collected in Fig.~\ref{fig:strain_vs_barrier_height}. The energy barriers for diffusion between adjacent A-sites range from $0.38\,\mathrm{eV}$ to $0.45\,\mathrm{eV}$ at $\xLi=0.25$, and from $0.27\,\mathrm{eV}$ to $0.48\,\mathrm{eV}$ at $\xLi=0.50$. These energies are broadly 
consistent with those from the previous DFT NEB study of Mees \etal \cite{MeesEtAl_PhysChemChemPhys2014}, and with values from NMR analysis of lithium jumps between lattice sites in these materials \cite{RoscianoEtAl_PhysChemChemPhys2013}. 

(Li,Al)-codoping of \chem{MgAl_2O_4} introduces disorder across the A-sites, and makes the lithium-occupied A-sites non-equivalent. From the perspective of a cation occupying a particular A-site, the local effect of this disorder is to introduce variation into the nearest-neighbour A-site occupation: each neighbouring A-site may now contain Li, Mg, or Al, or be vacant. For each lithium diffusion pathway, the nearest-neighbour A-site occupations for the end-point configurations allows paths to be classification as symmetric or asymmetric. Pathways with identical nearest-neighbour A-site coordination environments at both end-points are symmetric, while those with different coordination environments are asymmetric. 

For the symmetric pathways, the diffusion barrier height is necessarily equal for diffusion in both directions. For asymmetric pathways, however, the end-point energies may differ, and the potential energy profile relative to the starting structure depends on the direction of lithium diffusion. For the asymmetric pathways considered here, we present data for lithium diffusion in both forward and reverse directions, with the latter indicated with a superscript R. 

The relative energies of the two endpoints for an asymmetric path can be analysed in terms of $\sigma_\mathrm{A}$ (Eqn.~\ref{eqn:sigma_A}), which describes the average A-site oxidation-state deviation from that of \chem{MgAl_2O_4}. For the \lipath{}{U}{0.25}{3} path, the lower energy end-point gives $\sigma_\mathrm{A}=0.1069$, and the higher energy end-point gives $\sigma_\mathrm{A}=0.3207$. Similarly, for the \lipath{}{U}{0.50}{3} path, the lower energy end-point gives $\sigma_\mathrm{A}=0.2$, and the higher energy end-point gives $\sigma_\mathrm{A}=0.4721$. In both cases, the lower energy end-point gives the lower value of $\sigma_\mathrm{A}$, indicating a more uniform arrangement of formal charges on the spinel lattice.

The electrostatic potential profiles approximately follow the potential energy profiles. For each path, the calculated electrostatic potential barrier typically overestimates the NEB barrier. This is consistent with lithium ions experiencing stronger short-ranged repulsion at the tetrahedral A sites than at the intermediate octahedral sites. This increases the energies of lithium at each end-point relative to the NEB path maximum. For the \lipath{}{U}{0.5}{3} (and \lipath{}{U}{0.5}{3R}) paths, there is quantitative disagreement between the electrostatic potential profile and the NEB potential profile. The electrostatic potential predicts the incorrect energy ordering of the two end-points. This illustrates that the electrostatic potential alone is not guaranteed to give a good description of the potential energy surface for mobile lithium ions.

%********** Isotropic **********************

\subsection{Isotropically-strained (Li,Al)--co-doped spinel} 

\begin{figure*}[tb]
  \centering
  \resizebox{16cm}{!}{\includegraphics*{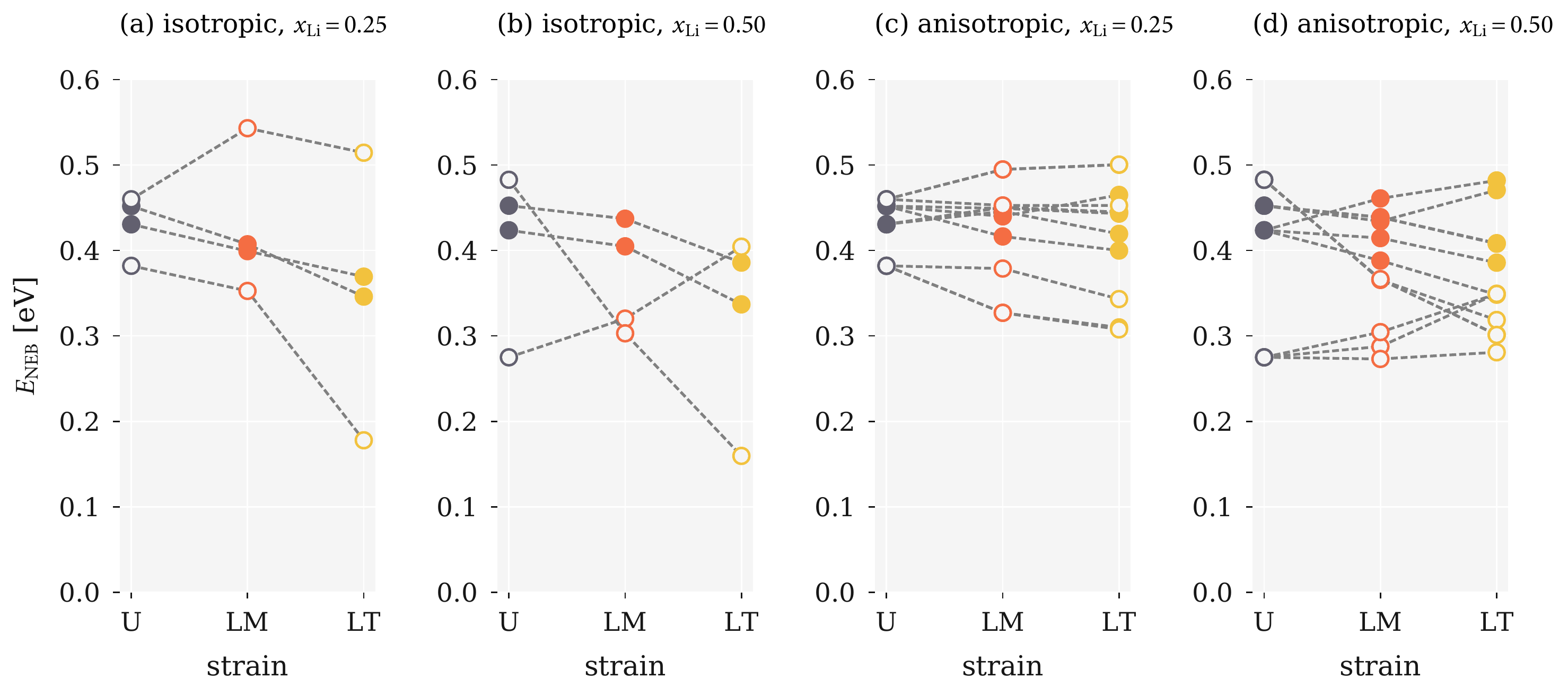}} %
    \caption{\label{fig:strain_vs_barrier_height}Variation in NEB barrier heights, $E_\mathrm{NEB}$, with strain. U = unstrained, LM = strained to \lmo{} lattice parameters, LT = strained to \lto{} lattice parameters. Panels (a) and (b) show data for isotropic strain, at $\xLi=0.25$ and $\xLi=0.50$, respectively. Panels (c) and (d) show data for anisotropic strain, at $\xLi=0.25$ and $\xLi=0.50$, respectively. Closed circles correspond to data for symmetric paths, and open circles to data for asymmetric paths.}
\end{figure*}

We now consider the effect of isotropic strain on lithium diffusion. We have calculated NEB barriers for both dopant concentrations with the unit cell strained to the 
\lmo{} and \lto{}
lattice volumes (Figs.~\ref{fig:neb_pathways_iso} and \ref{fig:strain_vs_barrier_height}). 
Isotropic strain has a large impact on the diffusion barriers. 
For the symmetric pathways the effect of this strain is straightforward: increased tensile strain decreases the diffusion barriers at both dopant concentrations. This result mirrors the effects seen in many solid oxide fuel cell electrolytes \cite{WenEtAl_JMaterChemA2015,Yildiz_MRSBull2014,AydinEtAl_PhysChemChemPhys2013,FluriEtAl_NatureComm2016}, where tensile strain typically increases oxide ion diffusion. 

\begin{figure}[tb]
  \centering
  \resizebox{7.2cm}{!}{\includegraphics*{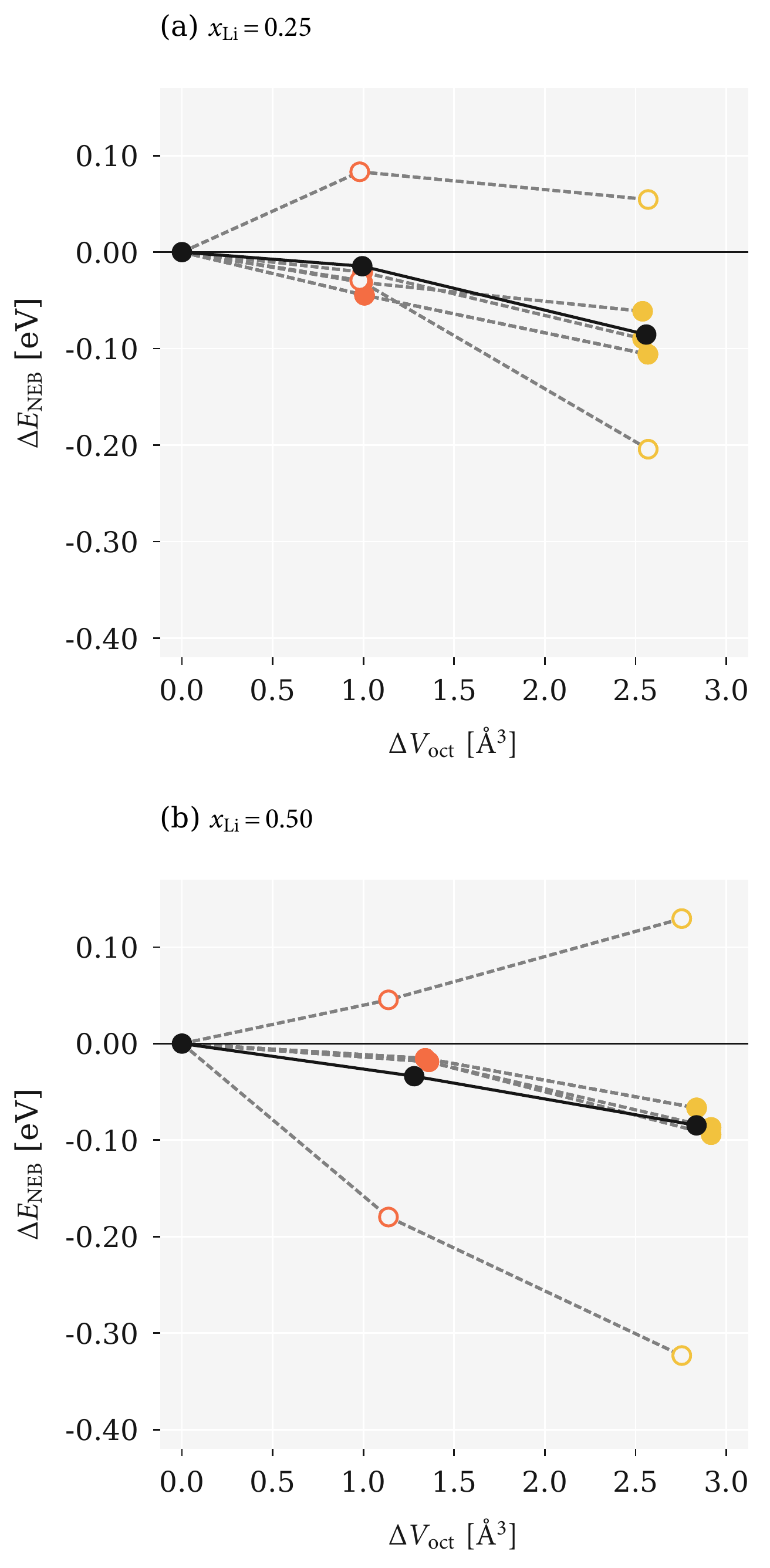}} %
    \caption{\label{fig:oct_vol_vs_delta_e_iso} Change in NEB barrier height,  $\Delta E_\mathrm{NEB}$, as a function of volume change for the 16c octahedra, $\Delta V_\mathrm{oct}$, under isotropic strain, for (a) $\xLi=0.25$ (b) $\xLi=0.5$. Red and orange circles correspond to strain to \lmo{} and \lto{} lattice parameters, respectively. Dashed lines link equivalent pathways under different strains. The solid black points show mean values under each strain condition.  Closed circles correspond to data for symmetric paths, and open circles to data for asymmetric paths.}
\end{figure}

For the asymmetric pathway the situation is more complicated. Because asymmetric paths necessarily have inequivalent endpoints, the applied strain can affect their \emph{relative} energies. The change in the diffusion barrier height therefore depends on the direction of motion along the path. At $\xLi=0.25$ the barrier is reduced for the forward pathway (3), but slightly increases for the reverse path (3R). At $\xLi=0.5$ this asymmetry is even more pronounced, with a large barrier increase for the forward path (3), but a large decrease for the reverse path (3R). Fig.~\ref{fig:oct_vol_vs_delta_e_iso} plots the relative change in volume for the intermediate octahedral site, $\Delta V_\mathrm{oct}$, against the change in barrier height, $\Delta E_\mathrm{NEB}$. For both $\xLi=0.25$ and $\xLi=0.50$, the average trend of decreasing the NEB barrier height is apparent. The outliers (showing both increased and strongly decreased barrier heights) correspond to the two asymmetric paths described above.

\begin{figure}[tb]
  \centering
  \resizebox{7.2cm}{!}{\includegraphics*{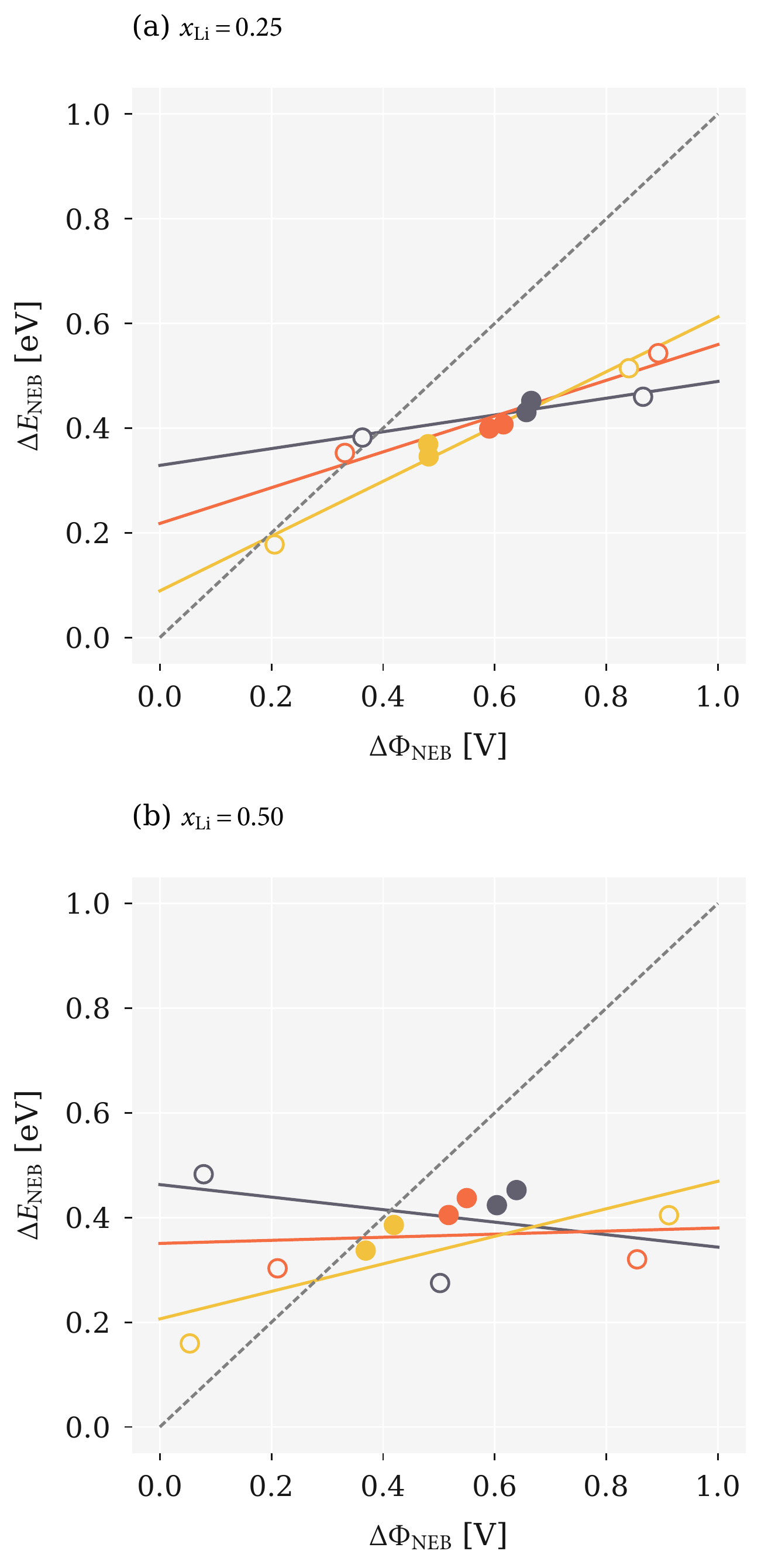}} %
    \caption{\label{fig:pot_vs_delta_e} Electrostatic potential barrier, $\Delta \Phi_\mathrm{NEB}$, versus potential energy barrier, $\Delta E_\mathrm{NEB}$, for the diffusing lithium ion, for pathways under zero-strain (black), and isotropically strained to the $\mathrm{Li}_y\mathrm{Mn}_2\mathrm{O}_4$ (red) and $\mathrm{Li}_{4+3z}\mathrm{Ti}_{5}\mathrm{O}_{12}$ (orange) cell volumes. (a) $\xLi=0.25$. (b) $\xLi=0.50$. The solid lines show linear best fits at each strain (U, LM, LT). The diagonal dashed line corresponds to an exact 1:1 relationship between $\Delta \Phi_\mathrm{NEB}$ and $\Delta E_\mathrm{NEB}$.  Closed circles correspond to data for symmetric paths, and open circles to data for asymmetric paths.}
\end{figure}

The effect of strain on the NEB profiles is mirrored by the electrostatic potential profiles, suggesting that changes in electrostatic potential do provide at least a qualitative metric for predicting the effects of strain on lithium ion diffusion. Fig.~\ref{fig:pot_vs_delta_e} plots the NEB barrier against the electrostatic barrier for each path. Linear--least-squares fits for each strain protocol show that the electrostatic potential barrier becomes an increasingly good predictor of the NEB barrier at larger strains. This is consistent with short-ranged repulsion interactions being less significant as the lattice volume increases under tensile strain.

The effect of isotropic strain on the asymmetric paths can also be examined from the perspective of the $\sigma_\mathrm{A}$ values for each end-point. For both asymmetric paths, $S_{0.25}^3$ and $S_{0.50}^3$, tensile strain produces a relative stabilisation of the end-point structure that more strongly deviates from the average +II oxidation state of the parent \chem{MgAl_2O_4} spinel, i.e.\ the structure with the larger $\sigma_\mathrm{A}$. At $\xLi=0.5$, this effect is large enough that the relative stabilities of the two end-points are reversed, contradicting the general trend that higher $\sigma$ values correspond to less stable structures. 

\subsection{Anisotropically-strained (Li,Al)--co-doped spinel}  

\begin{figure*}[tb]
  \centering
  \resizebox{18cm}{!}{\includegraphics*{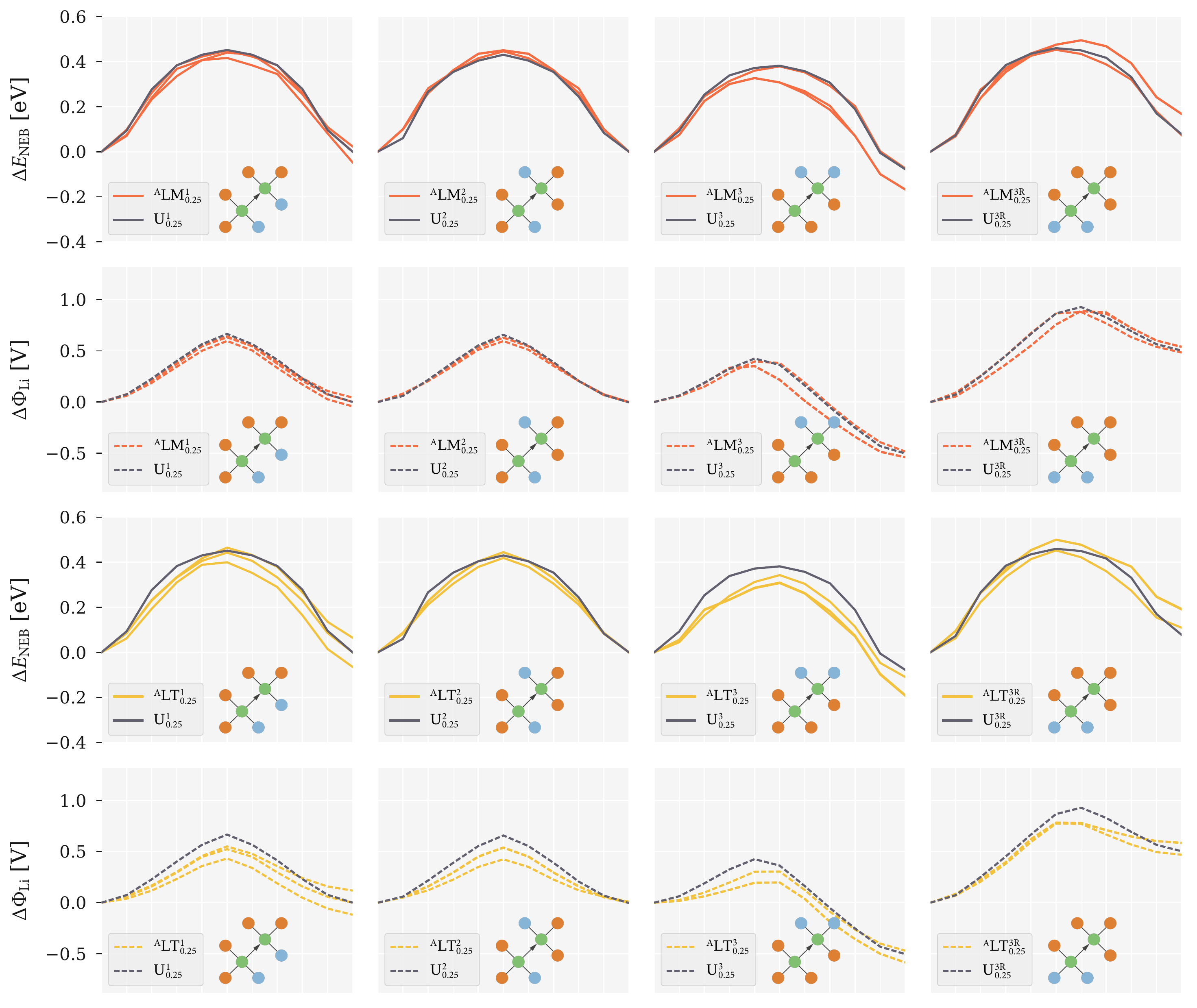}} %
    \caption{\label{fig:neb_pathways_aniso_025}$\xLi=0.25$ cNEB energy profiles and corresponding electrostatic potential profiles for unstrained and anisotropically strained structures. Notation is as follows: $S_x^p$, where $S$ is strain (U = unstrained, LM = strained to lithium manganate lattice parameters (top two rows), LT = strained to lithium titanate lattice parameters (bottom two rows)), $x$ is the dopant concentration and $p$ enumerates the pathways (where R refers to the same pathway in reverse). A superscript A indicates anisotropic strain along only two axes. The inset schematic in each panel shows the coordination environments of the initial and final A-site positions. Li ions are green, Mg are orange and Al are blue. Electrostatic potential profiles (including the contribution from the mobile Li$^+$ ion) at the Li$^+$ site along the diffusion pathway are shown by the corresponding coloured dashed lines.}
\end{figure*}

\begin{figure*}[tb]
  \centering
  \resizebox{18cm}{!}{\includegraphics*{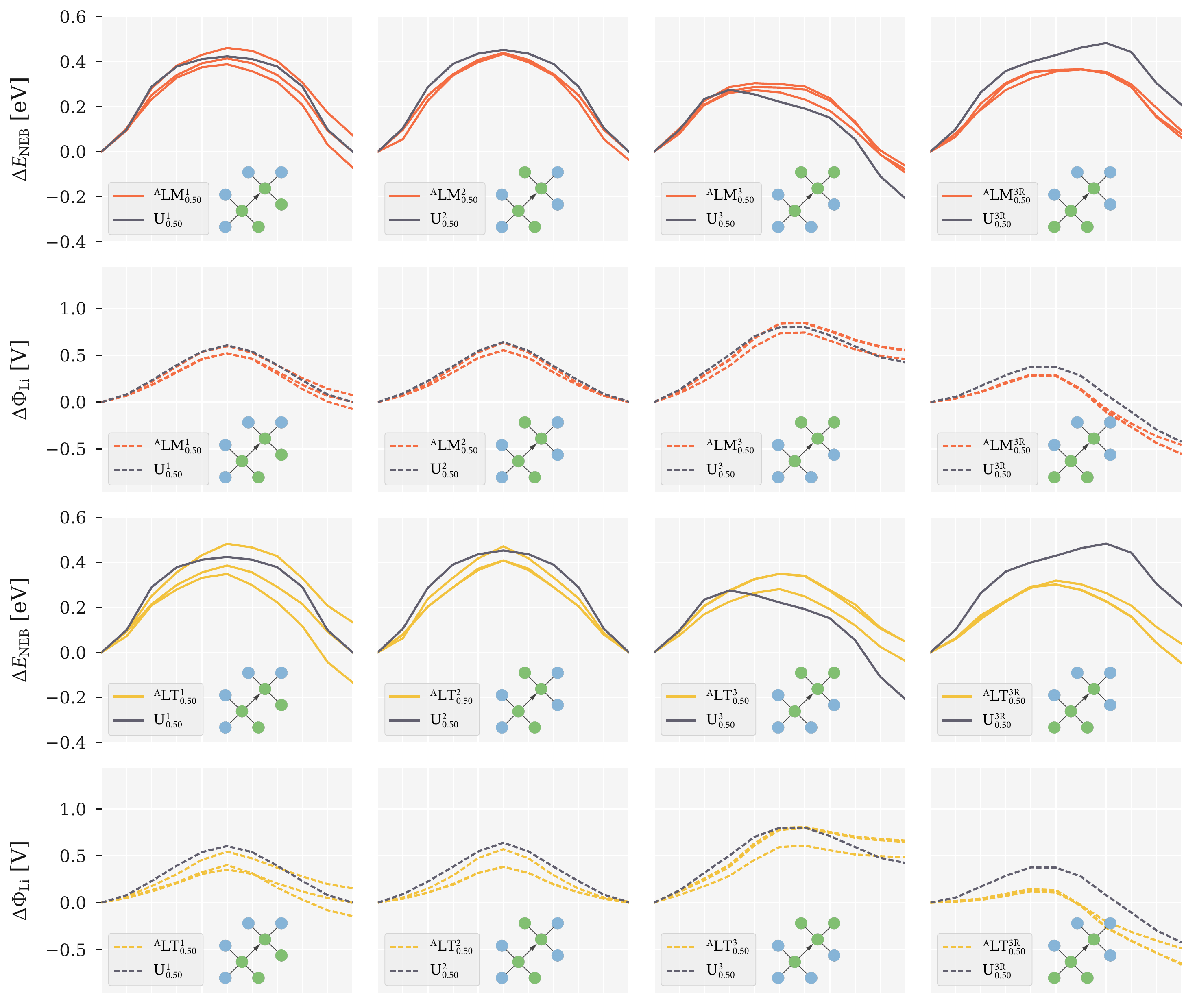}} %
    \caption{\label{fig:neb_pathways_aniso_050}$\xLi=0.50$ cNEB energy profiles and corresponding electrostatic potential profiles for unstrained and anisotropically strained structures. Notation is as follows: $S_x^p$, where $S$ is strain (U = unstrained, LM = strained to lithium manganate lattice parameters (top two rows), LT = strained to lithium titanate lattice parameters (bottom two rows)), $x$ is the dopant concentration and $p$ enumerates the pathways (where R refers to the same pathway in reverse). A superscript A indicates anisotropic strain along only two axes. The inset schematic in each panel shows the coordination environments of the initial and final A-site positions. Li ions are green, Mg are orange and Al are blue. Electrostatic potential profiles (including the contribution from the mobile Li$^+$ ion) at the Li$^+$ site along the diffusion pathway are shown by the corresponding coloured dashed lines.}
\end{figure*}

In the previous section, we have presented results showing the effect of isotropic strain on lithium-ion diffusion barriers. For a hypothetical lattice-matched electrolyte--electrode interface, however, lattice strain due to electrolyte--electrode epitaxy is not isotropic, but instead is anisotropic. Specifically, the lattice will be strained parallel to the interface, but is free to relax in the perpendicular direction. 

To model the strain at a hypothetical lattice-matched electrolyte--electrode interface, we extended our calculations on isotropically strained systems to consider the effect of anisotropic strain on the CI-NEB diffusion pathways. For these calculations, we still consider strain as arising from coherent $\{100\}_\mathrm{A}|\{100\}_\mathrm{B}$ heterointerfaces. 
Under anisotropic strain, the three $\left\{100\right\}$ directions become inequivalent, and we consider each possible strain orientation. For each calculation, the two in-plane lattice parameters are strained to match experimental values of either \lmo{} or \lto{}, and the third lattice parameter is optimised to minimise the total system energy. Because we consider only $\{100\}$ oriented interfaces, this is equivalent to biaxial strain. 

The effect of anisotropic strain on diffusion barrier height is much smaller than in the isotropic case, with changes from $-0.182\,\mathrm{eV}$ to $+0.074\,\mathrm{eV}$. Individual values depend on the exact path being considered, and on the relative orientation of the applied strain. Individual NEB profiles are shown in Fig.~\ref{fig:neb_pathways_aniso_025} ($\xLi=0.25$) and Fig.~\ref{fig:neb_pathways_aniso_050} ($\xLi=0.50$), and the full set of barrier heights are collected in Fig.~\ref{fig:strain_vs_barrier_height}.
The NEB pathway again depends on the symmetry of the start- and end-point A cation sites. In some cases, however, the application of planar strain breaks the symmetry found in the isotropic and unstrained structures. This point is illustrated by the \lipath{A}{LM}{0.25}{1} and \lipath{A}{LT}{0.25}{1} pathways (Fig.~\ref{fig:neb_pathways_aniso_025}) where the energy difference between the start- and end-points depends on the orientation of the applied strain.

\begin{figure}[tb]
  \centering
  \resizebox{7.2cm}{!}{\includegraphics*{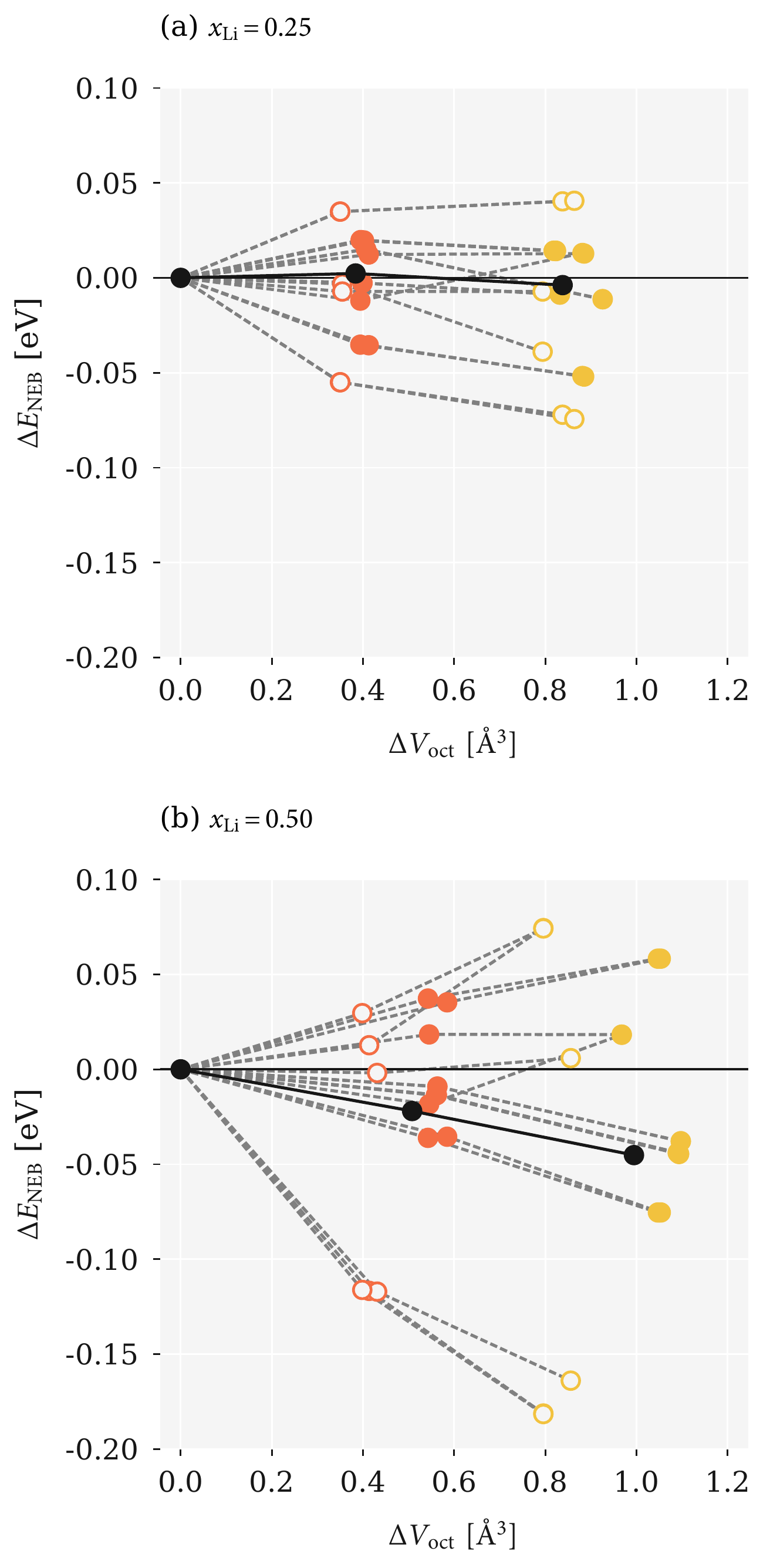}} %
    \caption{\label{fig:oct_vol_vs_delta_e_aniso}Change in NEB barrier height, $\Delta E_\mathrm{NEB}$, as a function of volume change for the 16c octahedra, $\Delta V_\mathrm{oct}$, under anisotropic strain, for (a) $\xLi=0.25$ (b) $\xLi=0.5$. Red and orange circles show strain to \lmo{} and \lto{} lattice parameters, respectively. Dashed lines link equivalent pathways under different strains. The solid black points show mean values under each strain condition. Closed circles correspond to data for symmetric paths, and open circles to data for asymmetric paths.}
\end{figure}

For the symmetric isotropically strained pathways, discussed above, the effect on the electrostatic potential under strain gave a good indication of the effect on the NEB diffusion barrier. This simple electrostatic model is not applicable, however, for the antisymmetric pathways. Under anisotropic strain, we find that the effect on the electrostatic potential is no longer a useful predictor of the change in NEB barrier: in some cases the electrostatic potential barrier decreases while the NEB barrier increases. The electrostatic potential profiles also show the effect of the applied strain lowering the cell symmetry, with relative potentials of path end-points now varying with the orientation of the applied strain. For isotropic strain, we observed a direct relationship between the volume of the transitional octahedral 16c site, and the electrostatic potential at the diffusing \chem{Li^+} site. For these anisotropically strained calculations, we find that the octahedral site volume changes are much smaller than those due to the equivalent isotropic strain, because the cell is allowed to contract along the free axis. As a result of this reduced volume change, the effect on diffusion barriers typically is much smaller than that of the isotropically strained equivalents (Figs.\ \ref{fig:neb_pathways_aniso_025},  \ref{fig:neb_pathways_aniso_050}, \& \ref{fig:oct_vol_vs_delta_e_aniso}).

\section{Summary \& Discussion}

(Li,Al)-codoped \chem{MgAl_2O_4} represents a new class of spinel-structured lithium-ion solid electrolytes, with potential applications in an all-spinel solid-state lithium-ion battery \cite{RoscianoEtAl_PhysChemChemPhys2013}, or as an electrode--buffer-layer \cite{PutEtAl_PhysChemChemPhys2015}, with lattice-matched electrolyte--electrode interfaces. Any lattice-matched (or epitaxial) interface between an electrode and electrolyte will exhibit some residual intrinsic strain. In the case of (Li,Al)-codoped \chem{MgAl_2O_4}, hypothetical lattice-matched $(100)$ interfaces with the spinel electrodes \lmo{} or \lto{} require tensile strains of up to $3.2$ or $6.0\,\%$ respectively, at $\xLi=0.5$. Because the lattice parameter of (Li,Al)-codoped \chem{MgAl_2O_4} decreases with higher doping levels, It has previously been suggested that high dopant concentrations ($\xLi>0.3$) should be avoided, to minimise possible corresponding increases in interfacial resistance \cite{MeesEtAl_PhysChemChemPhys2014}. High dopant concentrations, however, have been predicted to have higher lithium conductivities, making them more appealing as solid electrolytes. These issues have prompted us to conduct this computational study into the effects of tensile strain on the lithium-ion diffusion pathways in (Li,Al)-doped \chem{MgAl_2O_4}, to assess the extent to which the increased strain expected at higher dopant concentrations affects lithium-ion transport.

We have performed a series of climbing-image nudged--elastic-band (CI-NEB) calculations to evaluate the potential energy profile for lithium vacancy diffusion, which proceeds via tetrahedron--octahedron--tetrahedron paths through the spinel structure. Our calculations show that isotropic strain, for most paths, decreases diffusion barriers; with an average reduction of $\sim\!0.1\,\mathrm{eV}$ for a lattice strained to match \lto{} ($6\%$ strain). This change is correlated with changes in the electrostatic potential profile along each diffusion path. The electrostatic potential profile, however, does not give a good prediction of potential energy barriers for all paths, particularly in cases where paths are asymmetric. As the lattice volume increases under tensile strain, however, we find that the electrostatic potential profile gives an increasingly good approximation of the diffusion potential energy profile. 

For a realistic lattice-matched interface, the resulting strain will be anisotropic (epitaxial). From our CI-NEB calculations of anisotropically strained cells, in general we predict a much smaller change in diffusion barriers than for equivalent isotropic strain. Individual changes, however, show a complex dependency on the particular path being considered, and the orientation of the diffusion pathway relative to the applied strain, and some barriers increase, by up to $+0.074\,\mathrm{eV}$ at $6.0\%$ strain. The smaller effect on diffusion barriers under epitaxial strain compared to equivalent isotropic strain is attributed to the capacity of a system under epitaxial strain to relax in the perpendicular direction. This difference is can be seen in the relative volume changes of the intermediate octahedral sites. Under anisotropic strain the transitional octahedral sites undergo a smaller volume expansion than under equivalent isotropic strain. 

Because the distribution of barrier heights is predicted to not change significantly under anisotropic strain, particularly at $\xLi=0.5$, we postulate that epitaxial strain arising from lattice-matching will not significantly affect lithium diffusion in the region close to an electrode--electrolyte interface. The effect of epitaxial strain on individual barriers is complex, however, causing both small decreases and small increases in barrier heights. A precise quantitative prediction of the effect of strain on ensemble transport properties (diffusion coefficients and ionic conductivities), would therefore require statistically sampling all relevant paths, using, for example, molecular dynamics or kinetic Monte Carlo simulations (cf. ref.~\cite{MeesEtAl_PhysChemChemPhys2014}).

Changes to local diffusion barriers are not the only possible mechanism by which interfacial strain might affect interfacial resistance. At a heterointerface between an electrode and electrolyte, the standard chemical potential of lithium ions may differ between the two materials. This can drive a spontaneous redistribution of the mobile ions across the interface to form space-charge layers on each side \cite{MorganAndMadden_PhysRevB2014}. Under an applied voltage, further redistribution of ions can occur, producing more complex space-charge profiles \cite{LeungAndLeenheer_JPhysChemC2015}. The change in local lithium-ion concentration associated with space-charge formation can affect the local conductivity (and resistivity) either positively or negatively, and hence contributes to interfacial resistance. Because strain shifts the local electrostatic potential, interfacial strain  changes the standard electrochemical potential offset at an electrode--electrolyte interface. This consequently affects space-charge formation, and the associated space-charge contribution to interfacial resistance. Our study also implicitly assumes that any relevant epitaxial heterointerface between electrode and electrolyte can be formed without dislocations or other extended defects. A complete model of lattice-matched  electrode--electrolyte interfaces should, therefore, also consider these issues. Assuming a coherent lattice-matched interface, however, and neglecting shifts in electrostatic potentials, our results suggest that changes in local diffusion barriers due to epitaxial strain are small, even for relatively large in-plane strains. Extrapolating from this result, we posit that lithium transport close to lattice-matched electrolyte--electrode interfaces is not significantly affected by the residual lattice strain. Accurately resolving the transport behaviour for a specific electrode--electrolyte pair, however, requires going beyond the local diffusion barrier modelling described here, and explicitly calculating lithium transport coefficients under the appropriate strain.

\section{Data Access Statement} 
The DFT dataset supporting this study is available from the University of Bath Research Data Archive (doi: \href{https://dx.doi.org/10.15125/BATH-00438}{10.15125/BATH-00438}) \cite{ORourkeAndMorgan_SpinelNEBDataset2017}, published under the CC-BY-SA-4.0 license. This dataset contains input parameters and output files for all VASP calculations, and a series of Python scripts for collating relevant data from the VASP outputs. A Jupyter notebook containing code to produce Figs.~\ref{fig:neb_pathways_iso}--\ref{fig:oct_vol_vs_delta_e_aniso} is also available 
(Ref~\onlinecite{ORourkeAndMorgan_SpinelAnalysisNotebook2017}, doi: \href{https://dx.doi.org/10.5281/zenodo.1069417}{10.5281/zenodo.1069417}), published under the MIT license.

\section*{Acknowledgements}
This work was funded by EPSRC grant \href{http://gow.epsrc.ac.uk/NGBOViewGrant.aspx?GrantRef=EP/N004302/1}{EP/N004302/1}. 
B.\ J.\ M.\ acknowledges support from the Royal Society (UF130329). 
Calculations were performed using the Balena High Performance Computing Service at the University of Bath, and using the ARCHER supercomputer, with access through membership of the UK's HPC Materials Chemistry Consortium, funded by EPSRC grant \href{http://gow.epsrc.ac.uk/NGBOViewGrant.aspx?GrantRef=EP/L000202/1}{EP/L000202}.

\vspace{30mm}
\bibliography{Spinel}

\begin{thebibliography}{60}%
\makeatletter
\providecommand \@ifxundefined [1]{%
 \@ifx{#1\undefined}
}%
\providecommand \@ifnum [1]{%
 \ifnum #1\expandafter \@firstoftwo
 \else \expandafter \@secondoftwo
 \fi
}%
\providecommand \@ifx [1]{%
 \ifx #1\expandafter \@firstoftwo
 \else \expandafter \@secondoftwo
 \fi
}%
\providecommand \natexlab [1]{#1}%
\providecommand \enquote  [1]{``#1''}%
\providecommand \bibnamefont  [1]{#1}%
\providecommand \bibfnamefont [1]{#1}%
\providecommand \citenamefont [1]{#1}%
\providecommand \href@noop [0]{\@secondoftwo}%
\providecommand \href [0]{\begingroup \@sanitize@url \@href}%
\providecommand \@href[1]{\@@startlink{#1}\@@href}%
\providecommand \@@href[1]{\endgroup#1\@@endlink}%
\providecommand \@sanitize@url [0]{\catcode `\\12\catcode `\$12\catcode
  `\&12\catcode `\#12\catcode `\^12\catcode `\_12\catcode `\%12\relax}%
\providecommand \@@startlink[1]{}%
\providecommand \@@endlink[0]{}%
\providecommand \url  [0]{\begingroup\@sanitize@url \@url }%
\providecommand \@url [1]{\endgroup\@href {#1}{\urlprefix }}%
\providecommand \urlprefix  [0]{URL }%
\providecommand \Eprint [0]{\href }%
\providecommand \doibase [0]{http://dx.doi.org/}%
\providecommand \selectlanguage [0]{\@gobble}%
\providecommand \bibinfo  [0]{\@secondoftwo}%
\providecommand \bibfield  [0]{\@secondoftwo}%
\providecommand \translation [1]{[#1]}%
\providecommand \BibitemOpen [0]{}%
\providecommand \bibitemStop [0]{}%
\providecommand \bibitemNoStop [0]{.\EOS\space}%
\providecommand \EOS [0]{\spacefactor3000\relax}%
\providecommand \BibitemShut  [1]{\csname bibitem#1\endcsname}%
\let\auto@bib@innerbib\@empty
%</preamble>
\bibitem [{\citenamefont {Knauth}(2009)}]{Knauth_SolStatIonics2009}%
  \BibitemOpen
  \bibfield  {author} {\bibinfo {author} {\bibfnamefont {Philippe}\
  \bibnamefont {Knauth}},\ }\bibfield  {title} {\enquote {\bibinfo {title}
  {Inorganic solid {L}i ion conductors: An overview},}\ }\href {\doibase
  10.1016/j.ssi.2009.03.022} {\bibfield  {journal} {\bibinfo  {journal} {Sol.
  Stat. Ionics}\ }\textbf {\bibinfo {volume} {180}},\ \bibinfo {pages}
  {911--916} (\bibinfo {year} {2009})}\BibitemShut {NoStop}%
\bibitem [{\citenamefont {Bachman}\ \emph {et~al.}(2016)\citenamefont
  {Bachman}, \citenamefont {Muy}, \citenamefont {Grimaud}, \citenamefont
  {Chang}, \citenamefont {Pour}, \citenamefont {Lux}, \citenamefont {Paschos},
  \citenamefont {Maglia}, \citenamefont {Lupart}, \citenamefont {Lamp},
  \citenamefont {Giordano},\ and\ \citenamefont
  {Shao-Horn}}]{BachmanEtAl_ChemRev2016}%
  \BibitemOpen
  \bibfield  {author} {\bibinfo {author} {\bibfnamefont {John~Christopher}\
  \bibnamefont {Bachman}}, \bibinfo {author} {\bibfnamefont {Sokseiha}\
  \bibnamefont {Muy}}, \bibinfo {author} {\bibfnamefont {Alexis}\ \bibnamefont
  {Grimaud}}, \bibinfo {author} {\bibfnamefont {Hao-Hsun}\ \bibnamefont
  {Chang}}, \bibinfo {author} {\bibfnamefont {Nir}\ \bibnamefont {Pour}},
  \bibinfo {author} {\bibfnamefont {Simon~F.}\ \bibnamefont {Lux}}, \bibinfo
  {author} {\bibfnamefont {Odysseas}\ \bibnamefont {Paschos}}, \bibinfo
  {author} {\bibfnamefont {Filippo}\ \bibnamefont {Maglia}}, \bibinfo {author}
  {\bibfnamefont {Saskia}\ \bibnamefont {Lupart}}, \bibinfo {author}
  {\bibfnamefont {Peter}\ \bibnamefont {Lamp}}, \bibinfo {author}
  {\bibfnamefont {Livia}\ \bibnamefont {Giordano}}, \ and\ \bibinfo {author}
  {\bibfnamefont {Yang}\ \bibnamefont {Shao-Horn}},\ }\bibfield  {title}
  {\enquote {\bibinfo {title} {Inorganic solid-state electrolytes for lithium
  batteries: Mechanisms and properties governing ion conduction},}\ }\href
  {\doibase 10.1021/acs.chemrev.5b00563} {\bibfield  {journal} {\bibinfo
  {journal} {Chem. Rev.}\ }\textbf {\bibinfo {volume} {116}},\ \bibinfo {pages}
  {140--162} (\bibinfo {year} {2016})}\BibitemShut {NoStop}%
\bibitem [{\citenamefont {Manthiram}\ \emph {et~al.}(2017)\citenamefont
  {Manthiram}, \citenamefont {Yu},\ and\ \citenamefont
  {Wang}}]{ManthiramEtAl_NatRevMater2017}%
  \BibitemOpen
  \bibfield  {author} {\bibinfo {author} {\bibfnamefont {Arumugam}\
  \bibnamefont {Manthiram}}, \bibinfo {author} {\bibfnamefont {Xingwen}\
  \bibnamefont {Yu}}, \ and\ \bibinfo {author} {\bibfnamefont {Shaofei}\
  \bibnamefont {Wang}},\ }\bibfield  {title} {\enquote {\bibinfo {title}
  {Lithium battery chemistries enabled by solid-state electrolytes},}\
  }\href@noop {} {\bibfield  {journal} {\bibinfo  {journal} {Nat. Rev. Mater.}\
  }\textbf {\bibinfo {volume} {2}},\ \bibinfo {pages} {1--16} (\bibinfo {year}
  {2017})}\BibitemShut {NoStop}%
\bibitem [{\citenamefont {Kamaya}\ \emph {et~al.}(2011)\citenamefont {Kamaya},
  \citenamefont {Homma}, \citenamefont {Yamakawa}, \citenamefont {Hirayama},
  \citenamefont {Kanno}, \citenamefont {Yonemura}, \citenamefont {Kamiyama},
  \citenamefont {Kato}, \citenamefont {Hama}, \citenamefont {Kawamoto},\ and\
  \citenamefont {Mitsui}}]{KamayaEtAl_NatMater2011}%
  \BibitemOpen
  \bibfield  {author} {\bibinfo {author} {\bibfnamefont {Noriaki}\ \bibnamefont
  {Kamaya}}, \bibinfo {author} {\bibfnamefont {Kenji}\ \bibnamefont {Homma}},
  \bibinfo {author} {\bibfnamefont {Yuichiro}\ \bibnamefont {Yamakawa}},
  \bibinfo {author} {\bibfnamefont {Masaaki}\ \bibnamefont {Hirayama}},
  \bibinfo {author} {\bibfnamefont {Ryoji}\ \bibnamefont {Kanno}}, \bibinfo
  {author} {\bibfnamefont {Masao}\ \bibnamefont {Yonemura}}, \bibinfo {author}
  {\bibfnamefont {Takashi}\ \bibnamefont {Kamiyama}}, \bibinfo {author}
  {\bibfnamefont {Yuki}\ \bibnamefont {Kato}}, \bibinfo {author} {\bibfnamefont
  {Shigenori}\ \bibnamefont {Hama}}, \bibinfo {author} {\bibfnamefont {Koji}\
  \bibnamefont {Kawamoto}}, \ and\ \bibinfo {author} {\bibfnamefont {Akio}\
  \bibnamefont {Mitsui}},\ }\bibfield  {title} {\enquote {\bibinfo {title} {A
  lithium superionic conductor},}\ }\href@noop {} {\bibfield  {journal}
  {\bibinfo  {journal} {Nat. Mater.}\ }\textbf {\bibinfo {volume} {10}},\
  \bibinfo {pages} {682--686} (\bibinfo {year} {2011})}\BibitemShut {NoStop}%
\bibitem [{\citenamefont {Santhanagopalan}\ \emph {et~al.}(2014)\citenamefont
  {Santhanagopalan}, \citenamefont {Qian}, \citenamefont {McGilvray},
  \citenamefont {Wang}, \citenamefont {Wang}, \citenamefont {Camino},
  \citenamefont {Graetz}, \citenamefont {Dudney},\ and\ \citenamefont
  {Meng}}]{SanthanagopalanEtAl_JPhysChemLett2014}%
  \BibitemOpen
  \bibfield  {author} {\bibinfo {author} {\bibfnamefont {Dhamodaran}\
  \bibnamefont {Santhanagopalan}}, \bibinfo {author} {\bibfnamefont {Danna}\
  \bibnamefont {Qian}}, \bibinfo {author} {\bibfnamefont {Thomas}\ \bibnamefont
  {McGilvray}}, \bibinfo {author} {\bibfnamefont {Ziying}\ \bibnamefont
  {Wang}}, \bibinfo {author} {\bibfnamefont {Feng}\ \bibnamefont {Wang}},
  \bibinfo {author} {\bibfnamefont {Fernando}\ \bibnamefont {Camino}}, \bibinfo
  {author} {\bibfnamefont {Jason}\ \bibnamefont {Graetz}}, \bibinfo {author}
  {\bibfnamefont {Nancy}\ \bibnamefont {Dudney}}, \ and\ \bibinfo {author}
  {\bibfnamefont {Ying~Shirley}\ \bibnamefont {Meng}},\ }\bibfield  {title}
  {\enquote {\bibinfo {title} {Interface limited lithium transport in
  solid-state batteries},}\ }\href@noop {} {\bibfield  {journal} {\bibinfo
  {journal} {J. Phys. Chem. Lett.}\ }\textbf {\bibinfo {volume} {5}},\ \bibinfo
  {pages} {298--303} (\bibinfo {year} {2014})}\BibitemShut {NoStop}%
\bibitem [{\citenamefont {Rosciano}\ \emph {et~al.}(2013)\citenamefont
  {Rosciano}, \citenamefont {Pescarmona}, \citenamefont {Houthoofd},
  \citenamefont {Persoons}, \citenamefont {Bottke},\ and\ \citenamefont
  {Wilkening}}]{RoscianoEtAl_PhysChemChemPhys2013}%
  \BibitemOpen
  \bibfield  {author} {\bibinfo {author} {\bibfnamefont {Fabio}\ \bibnamefont
  {Rosciano}}, \bibinfo {author} {\bibfnamefont {Paolo~P.}\ \bibnamefont
  {Pescarmona}}, \bibinfo {author} {\bibfnamefont {Kristof}\ \bibnamefont
  {Houthoofd}}, \bibinfo {author} {\bibfnamefont {Andre}\ \bibnamefont
  {Persoons}}, \bibinfo {author} {\bibfnamefont {Patrick}\ \bibnamefont
  {Bottke}}, \ and\ \bibinfo {author} {\bibfnamefont {Martin}\ \bibnamefont
  {Wilkening}},\ }\bibfield  {title} {\enquote {\bibinfo {title} {Towards a
  lattice-matching solid-state battery: synthesis of a new class of lithium-ion
  conductors with the spinel structure},}\ }\href@noop {} {\bibfield  {journal}
  {\bibinfo  {journal} {Phys. Chem. Chem. Phys.}\ }\textbf {\bibinfo {volume}
  {15}},\ \bibinfo {pages} {6107--6112} (\bibinfo {year} {2013})}\BibitemShut
  {NoStop}%
\bibitem [{\citenamefont {Thackeray}\ and\ \citenamefont
  {Goodenough}(1985)}]{Thackeray_Patent1985}%
  \BibitemOpen
  \bibfield  {author} {\bibinfo {author} {\bibfnamefont {M.~M.}\ \bibnamefont
  {Thackeray}}\ and\ \bibinfo {author} {\bibfnamefont {J.~B.}\ \bibnamefont
  {Goodenough}},\ }\href {http://www.google.co.uk/patents/US4507371} {\enquote
  {\bibinfo {title} {Solid state cell wherein an anode, solid electrolyte and
  cathode each comprise a cubic-close-packed framework structure},}\ }
  (\bibinfo {year} {1985}),\ \bibinfo {note} {{US} Patent
  4,507,371}\BibitemShut {NoStop}%
\bibitem [{\citenamefont {Blaakmeer}\ \emph {et~al.}(2015)\citenamefont
  {Blaakmeer}, \citenamefont {Rosciano},\ and\ \citenamefont {van
  Eck}}]{BlaakmeerEtAl_JPhysChemC2015}%
  \BibitemOpen
  \bibfield  {author} {\bibinfo {author} {\bibfnamefont {E.~S.~(Merijn)}\
  \bibnamefont {Blaakmeer}}, \bibinfo {author} {\bibfnamefont {Fabio}\
  \bibnamefont {Rosciano}}, \ and\ \bibinfo {author} {\bibfnamefont {Ernst
  R.~H.}\ \bibnamefont {van Eck}},\ }\bibfield  {title} {\enquote {\bibinfo
  {title} {Lithium doping of {M}g{A}l$_2${O}$_4$ and {Z}n{A}l$_2${O}$_4$
  investigated by high-resolution solid state {NMR}},}\ }\href {\doibase
  10.1021/jp512304e} {\bibfield  {journal} {\bibinfo  {journal} {J. Phys. Chem.
  C}\ }\textbf {\bibinfo {volume} {119}},\ \bibinfo {pages} {7565--7577}
  (\bibinfo {year} {2015})}\BibitemShut {NoStop}%
\bibitem [{\citenamefont {Djenadic}\ \emph {et~al.}(2016)\citenamefont
  {Djenadic}, \citenamefont {Botros},\ and\ \citenamefont
  {Hahn}}]{DjenadicEtAl_SolStatIonics2016}%
  \BibitemOpen
  \bibfield  {author} {\bibinfo {author} {\bibfnamefont {Ruzica}\ \bibnamefont
  {Djenadic}}, \bibinfo {author} {\bibfnamefont {Miriam}\ \bibnamefont
  {Botros}}, \ and\ \bibinfo {author} {\bibfnamefont {Horst}\ \bibnamefont
  {Hahn}},\ }\bibfield  {title} {\enquote {\bibinfo {title} {Is {L}i-doped
  {M}g{A}l$_{2}${O}$_{4}$ a potential solid electrolyte for an all-spinel
  {L}i-ion battery?}}\ }\href@noop {} {\bibfield  {journal} {\bibinfo
  {journal} {Sol. Stat. Ionics}\ }\textbf {\bibinfo {volume} {287}},\ \bibinfo
  {pages} {71--76} (\bibinfo {year} {2016})}\BibitemShut {NoStop}%
\bibitem [{\citenamefont {Yi}\ \emph {et~al.}(2009)\citenamefont {Yi},
  \citenamefont {Zhu}, \citenamefont {Zhu}, \citenamefont {Shu}, \citenamefont
  {Yue},\ and\ \citenamefont {Zhou}}]{YiEtAl_Ionics2009}%
  \BibitemOpen
  \bibfield  {author} {\bibinfo {author} {\bibfnamefont {Ting-Feng}\
  \bibnamefont {Yi}}, \bibinfo {author} {\bibfnamefont {Yan-Rong}\ \bibnamefont
  {Zhu}}, \bibinfo {author} {\bibfnamefont {Xiao-Dong}\ \bibnamefont {Zhu}},
  \bibinfo {author} {\bibfnamefont {J.}~\bibnamefont {Shu}}, \bibinfo {author}
  {\bibfnamefont {Cai-Bo}\ \bibnamefont {Yue}}, \ and\ \bibinfo {author}
  {\bibfnamefont {An-Na}\ \bibnamefont {Zhou}},\ }\bibfield  {title} {\enquote
  {\bibinfo {title} {A review of recent developments in the surface
  modification of {L}i{M}n$_2${O}$_4$ as cathode material of power lithium-ion
  battery},}\ }\href@noop {} {\bibfield  {journal} {\bibinfo  {journal}
  {Ionics}\ }\textbf {\bibinfo {volume} {15}},\ \bibinfo {pages} {779--784}
  (\bibinfo {year} {2009})}\BibitemShut {NoStop}%
\bibitem [{\citenamefont {Chen}\ \emph {et~al.}(2010)\citenamefont {Chen},
  \citenamefont {Qin}, \citenamefont {Amine},\ and\ \citenamefont
  {Sun}}]{ChenEtAl_JMaterChem2010}%
  \BibitemOpen
  \bibfield  {author} {\bibinfo {author} {\bibfnamefont {Zonghai}\ \bibnamefont
  {Chen}}, \bibinfo {author} {\bibfnamefont {Yan}\ \bibnamefont {Qin}},
  \bibinfo {author} {\bibfnamefont {Khalil}\ \bibnamefont {Amine}}, \ and\
  \bibinfo {author} {\bibfnamefont {Y.~K.}\ \bibnamefont {Sun}},\ }\bibfield
  {title} {\enquote {\bibinfo {title} {Role of surface coating on cathode
  materials for lithium-ion batteries},}\ }\href@noop {} {\bibfield  {journal}
  {\bibinfo  {journal} {J. Mater. Chem.}\ }\textbf {\bibinfo {volume} {20}},\
  \bibinfo {pages} {7606--7} (\bibinfo {year} {2010})}\BibitemShut {NoStop}%
\bibitem [{\citenamefont {Aykol}\ \emph {et~al.}(2016)\citenamefont {Aykol},
  \citenamefont {Kim}, \citenamefont {Hegde}, \citenamefont {Snydacker},
  \citenamefont {Lu}, \citenamefont {Hao}, \citenamefont {Kirklin},
  \citenamefont {Morgan},\ and\ \citenamefont
  {Wolverton}}]{AykolEtAl_NatureComm2016}%
  \BibitemOpen
  \bibfield  {author} {\bibinfo {author} {\bibfnamefont {Muratahan}\
  \bibnamefont {Aykol}}, \bibinfo {author} {\bibfnamefont {Soo}\ \bibnamefont
  {Kim}}, \bibinfo {author} {\bibfnamefont {Vinay~I.}\ \bibnamefont {Hegde}},
  \bibinfo {author} {\bibfnamefont {David}\ \bibnamefont {Snydacker}}, \bibinfo
  {author} {\bibfnamefont {Zhi}\ \bibnamefont {Lu}}, \bibinfo {author}
  {\bibfnamefont {Shiqiang}\ \bibnamefont {Hao}}, \bibinfo {author}
  {\bibfnamefont {Scott}\ \bibnamefont {Kirklin}}, \bibinfo {author}
  {\bibfnamefont {Dane}\ \bibnamefont {Morgan}}, \ and\ \bibinfo {author}
  {\bibfnamefont {C.}~\bibnamefont {Wolverton}},\ }\bibfield  {title} {\enquote
  {\bibinfo {title} {High-throughput computational design of cathode coatings
  for {L}i-ion batteries},}\ }\href@noop {} {\bibfield  {journal} {\bibinfo
  {journal} {Nature Comm.}\ }\textbf {\bibinfo {volume} {7}},\ \bibinfo {pages}
  {1--12} (\bibinfo {year} {2016})}\BibitemShut {NoStop}%
\bibitem [{\citenamefont {Zuo}\ \emph {et~al.}(2017)\citenamefont {Zuo},
  \citenamefont {Tian}, \citenamefont {Li}, \citenamefont {Chen},\ and\
  \citenamefont {Shu}}]{ZuoEtAl_JAllCom2017}%
  \BibitemOpen
  \bibfield  {author} {\bibinfo {author} {\bibfnamefont {Daxian}\ \bibnamefont
  {Zuo}}, \bibinfo {author} {\bibfnamefont {Guanglei}\ \bibnamefont {Tian}},
  \bibinfo {author} {\bibfnamefont {Xiang}\ \bibnamefont {Li}}, \bibinfo
  {author} {\bibfnamefont {Da}~\bibnamefont {Chen}}, \ and\ \bibinfo {author}
  {\bibfnamefont {Kangying}\ \bibnamefont {Shu}},\ }\bibfield  {title}
  {\enquote {\bibinfo {title} {Recent progress in surface coating of cathode
  materials for lithium ion secondary batteries},}\ }\href@noop {} {\bibfield
  {journal} {\bibinfo  {journal} {J. All. Com.}\ }\textbf {\bibinfo {volume}
  {706}},\ \bibinfo {pages} {24--40} (\bibinfo {year} {2017})}\BibitemShut
  {NoStop}%
\bibitem [{\citenamefont {Ohta}\ \emph {et~al.}(2006)\citenamefont {Ohta},
  \citenamefont {Takada}, \citenamefont {Zhang}, \citenamefont {Ma},
  \citenamefont {Osada},\ and\ \citenamefont {Sasaki}}]{OhtaEtAl_AdvMater2006}%
  \BibitemOpen
  \bibfield  {author} {\bibinfo {author} {\bibfnamefont {N.}~\bibnamefont
  {Ohta}}, \bibinfo {author} {\bibfnamefont {K.}~\bibnamefont {Takada}},
  \bibinfo {author} {\bibfnamefont {L.}~\bibnamefont {Zhang}}, \bibinfo
  {author} {\bibfnamefont {R.}~\bibnamefont {Ma}}, \bibinfo {author}
  {\bibfnamefont {M.}~\bibnamefont {Osada}}, \ and\ \bibinfo {author}
  {\bibfnamefont {T.}~\bibnamefont {Sasaki}},\ }\bibfield  {title} {\enquote
  {\bibinfo {title} {Enhancement of the high-rate capability of solid-state
  lithium batteries by nanoscale interfacial modification},}\ }\href@noop {}
  {\bibfield  {journal} {\bibinfo  {journal} {Adv. Mater.}\ }\textbf {\bibinfo
  {volume} {18}},\ \bibinfo {pages} {2226--2229} (\bibinfo {year}
  {2006})}\BibitemShut {NoStop}%
\bibitem [{\citenamefont {Ohta}\ \emph {et~al.}(2007)\citenamefont {Ohta},
  \citenamefont {Takada}, \citenamefont {Sakaguchi}, \citenamefont {Zhang},
  \citenamefont {Ma}, \citenamefont {Fukuda}, \citenamefont {Osada},\ and\
  \citenamefont {Sasaki}}]{OhtaEtAl_ElectrochemComm2007}%
  \BibitemOpen
  \bibfield  {author} {\bibinfo {author} {\bibfnamefont {Narumi}\ \bibnamefont
  {Ohta}}, \bibinfo {author} {\bibfnamefont {Kazunori}\ \bibnamefont {Takada}},
  \bibinfo {author} {\bibfnamefont {Isao}\ \bibnamefont {Sakaguchi}}, \bibinfo
  {author} {\bibfnamefont {Lianqi}\ \bibnamefont {Zhang}}, \bibinfo {author}
  {\bibfnamefont {Renzhi}\ \bibnamefont {Ma}}, \bibinfo {author} {\bibfnamefont
  {Katsutoshi}\ \bibnamefont {Fukuda}}, \bibinfo {author} {\bibfnamefont
  {Minoru}\ \bibnamefont {Osada}}, \ and\ \bibinfo {author} {\bibfnamefont
  {Takayoshi}\ \bibnamefont {Sasaki}},\ }\bibfield  {title} {\enquote {\bibinfo
  {title} {{L}i{N}b{O}$_3$-coated {L}i{C}o{O}$_2$ as cathode material for all
  solid-state lithium secondary batteries},}\ }\href@noop {} {\bibfield
  {journal} {\bibinfo  {journal} {Electrochem Comm.}\ }\textbf {\bibinfo
  {volume} {9}},\ \bibinfo {pages} {1486--1490} (\bibinfo {year}
  {2007})}\BibitemShut {NoStop}%
\bibitem [{\citenamefont {Takada}\ and\ \citenamefont
  {Ohno}(2016)}]{TakadaAndOhno_FrontEnergyRes2016}%
  \BibitemOpen
  \bibfield  {author} {\bibinfo {author} {\bibfnamefont {Kazunori}\
  \bibnamefont {Takada}}\ and\ \bibinfo {author} {\bibfnamefont {Takahisa}\
  \bibnamefont {Ohno}},\ }\bibfield  {title} {\enquote {\bibinfo {title}
  {Experimental and computational approaches to interfacial resistance in
  solid-state batteries},}\ }\href@noop {} {\bibfield  {journal} {\bibinfo
  {journal} {Front. Energy Res.}\ }\textbf {\bibinfo {volume} {4}},\ \bibinfo
  {pages} {014101--7} (\bibinfo {year} {2016})}\BibitemShut {NoStop}%
\bibitem [{\citenamefont {Li}\ \emph {et~al.}(2014)\citenamefont {Li},
  \citenamefont {Zhu}, \citenamefont {Wang},\ and\ \citenamefont
  {Cao}}]{LiEtAl_ACSApplMaterInt2014}%
  \BibitemOpen
  \bibfield  {author} {\bibinfo {author} {\bibfnamefont {Jili}\ \bibnamefont
  {Li}}, \bibinfo {author} {\bibfnamefont {Youqi}\ \bibnamefont {Zhu}},
  \bibinfo {author} {\bibfnamefont {Lin}\ \bibnamefont {Wang}}, \ and\ \bibinfo
  {author} {\bibfnamefont {Chuanbao}\ \bibnamefont {Cao}},\ }\bibfield  {title}
  {\enquote {\bibinfo {title} {Lithium titanate epitaxial coating on spinel
  lithium manganese oxide surface for improving the performance of lithium
  storage capability},}\ }\href@noop {} {\bibfield  {journal} {\bibinfo
  {journal} {ACS Appl. Mater. Int.}\ }\textbf {\bibinfo {volume} {6}},\
  \bibinfo {pages} {18742--18750} (\bibinfo {year} {2014})}\BibitemShut
  {NoStop}%
\bibitem [{\citenamefont {Berg}\ and\ \citenamefont
  {Thomas}(1999)}]{BergAndThomas_SolStatIonics1999}%
  \BibitemOpen
  \bibfield  {author} {\bibinfo {author} {\bibfnamefont {H.}~\bibnamefont
  {Berg}}\ and\ \bibinfo {author} {\bibfnamefont {J.~O.}\ \bibnamefont
  {Thomas}},\ }\bibfield  {title} {\enquote {\bibinfo {title} {Neutron
  diffraction study of electrochemically delithiated {L}i{M}n$_2${O}$_4$
  spinel},}\ }\href {\doibase http://dx.doi.org/10.1016/S0167-2738(99)00235-0}
  {\bibfield  {journal} {\bibinfo  {journal} {Sol. Stat. Ionics}\ }\textbf
  {\bibinfo {volume} {126}},\ \bibinfo {pages} {227 -- 234} (\bibinfo {year}
  {1999})}\BibitemShut {NoStop}%
\bibitem [{\citenamefont {Wagemaker}\ \emph {et~al.}(2006)\citenamefont
  {Wagemaker}, \citenamefont {Simon}, \citenamefont {Kelder}, \citenamefont
  {Schoonman}, \citenamefont {Ringpfeil}, \citenamefont {Haake}, \citenamefont
  {L{\"u}tzenkirchen-Hecht}, \citenamefont {Frahm},\ and\ \citenamefont
  {Mulder}}]{WagemakerEtAl_AdvMater2006}%
  \BibitemOpen
  \bibfield  {author} {\bibinfo {author} {\bibfnamefont {M.}~\bibnamefont
  {Wagemaker}}, \bibinfo {author} {\bibfnamefont {D. R.}\ \bibnamefont
  {Simon}}, \bibinfo {author} {\bibfnamefont {E. M.}\ \bibnamefont {Kelder}},
  \bibinfo {author} {\bibfnamefont {J.}~\bibnamefont {Schoonman}}, \bibinfo
  {author} {\bibfnamefont {C.}~\bibnamefont {Ringpfeil}}, \bibinfo {author}
  {\bibfnamefont {U.}~\bibnamefont {Haake}}, \bibinfo {author} {\bibfnamefont
  {D.}~\bibnamefont {L{\"u}tzenkirchen-Hecht}}, \bibinfo {author}
  {\bibfnamefont {R.}~\bibnamefont {Frahm}}, \ and\ \bibinfo {author}
  {\bibfnamefont {F. M.}\ \bibnamefont {Mulder}},\ }\bibfield  {title}
  {\enquote {\bibinfo {title} {A kinetic two-phase and equilibrium solid
  solution in spinel {L}i$_{4+x}${T}i$_5${O}$_{12}$},}\ }\href {\doibase
  10.1002/adma.200601636} {\bibfield  {journal} {\bibinfo  {journal} {Adv.
  Mater.}\ }\textbf {\bibinfo {volume} {18}},\ \bibinfo {pages} {3169--3173}
  (\bibinfo {year} {2006})}\BibitemShut {NoStop}%
\bibitem [{\citenamefont {Schichtel}\ \emph {et~al.}(2009)\citenamefont
  {Schichtel}, \citenamefont {Korte}, \citenamefont {Hesse},\ and\
  \citenamefont {Janek}}]{SchichtelEtAl_PhysChemChemPhys2009}%
  \BibitemOpen
  \bibfield  {author} {\bibinfo {author} {\bibfnamefont {N.}~\bibnamefont
  {Schichtel}}, \bibinfo {author} {\bibfnamefont {C.}~\bibnamefont {Korte}},
  \bibinfo {author} {\bibfnamefont {D.}~\bibnamefont {Hesse}}, \ and\ \bibinfo
  {author} {\bibfnamefont {J.}~\bibnamefont {Janek}},\ }\bibfield  {title}
  {\enquote {\bibinfo {title} {Elastic strain at interfaces and its influence
  on ionic conductivity in nanoscaled solid electrolyte thin
  films---theoretical considerations and experimental studies},}\ }\href@noop
  {} {\bibfield  {journal} {\bibinfo  {journal} {Phys. Chem. Chem. Phys.}\
  }\textbf {\bibinfo {volume} {11}},\ \bibinfo {pages} {3043} (\bibinfo {year}
  {2009})}\BibitemShut {NoStop}%
\bibitem [{\citenamefont {Rupp}(2012)}]{Rupp_SolStatIonics2012}%
  \BibitemOpen
  \bibfield  {author} {\bibinfo {author} {\bibfnamefont {Jennifer L.~M.}\
  \bibnamefont {Rupp}},\ }\bibfield  {title} {\enquote {\bibinfo {title} {Ionic
  diffusion as a matter of lattice-strain for electroceramic thin films},}\
  }\href@noop {} {\bibfield  {journal} {\bibinfo  {journal} {Sol. Stat.
  Ionics}\ }\textbf {\bibinfo {volume} {207}},\ \bibinfo {pages} {1--13}
  (\bibinfo {year} {2012})}\BibitemShut {NoStop}%
\bibitem [{\citenamefont {Wen}\ \emph {et~al.}(2015)\citenamefont {Wen},
  \citenamefont {Lv},\ and\ \citenamefont {He}}]{WenEtAl_JMaterChemA2015}%
  \BibitemOpen
  \bibfield  {author} {\bibinfo {author} {\bibfnamefont {Kechun}\ \bibnamefont
  {Wen}}, \bibinfo {author} {\bibfnamefont {Weiqiang}\ \bibnamefont {Lv}}, \
  and\ \bibinfo {author} {\bibfnamefont {Weidong}\ \bibnamefont {He}},\
  }\bibfield  {title} {\enquote {\bibinfo {title} {Interfacial lattice-strain
  effects on improving the overall performance of micro-solid oxide fuel
  cells},}\ }\href@noop {} {\bibfield  {journal} {\bibinfo  {journal} {J.
  Mater. Chem. A}\ }\textbf {\bibinfo {volume} {3}},\ \bibinfo {pages}
  {20031--20050} (\bibinfo {year} {2015})}\BibitemShut {NoStop}%
\bibitem [{\citenamefont {Yildiz}(2014)}]{Yildiz_MRSBull2014}%
  \BibitemOpen
  \bibfield  {author} {\bibinfo {author} {\bibfnamefont {Bilge}\ \bibnamefont
  {Yildiz}},\ }\bibfield  {title} {\enquote {\bibinfo {title} {{``Stretching''
  the energy landscape of oxides---Effects on electrocatalysis and
  diffusion}},}\ }\href@noop {} {\bibfield  {journal} {\bibinfo  {journal} {MRS
  Bull.}\ }\textbf {\bibinfo {volume} {39}},\ \bibinfo {pages} {147--156}
  (\bibinfo {year} {2014})}\BibitemShut {NoStop}%
\bibitem [{\citenamefont {Aydin}\ \emph {et~al.}(2013)\citenamefont {Aydin},
  \citenamefont {Korte}, \citenamefont {Rohnke},\ and\ \citenamefont
  {Janek}}]{AydinEtAl_PhysChemChemPhys2013}%
  \BibitemOpen
  \bibfield  {author} {\bibinfo {author} {\bibfnamefont {Halit}\ \bibnamefont
  {Aydin}}, \bibinfo {author} {\bibfnamefont {Carsten}\ \bibnamefont {Korte}},
  \bibinfo {author} {\bibfnamefont {Marcus}\ \bibnamefont {Rohnke}}, \ and\
  \bibinfo {author} {\bibfnamefont {J{\"u}rgen}\ \bibnamefont {Janek}},\
  }\bibfield  {title} {\enquote {\bibinfo {title} {Oxygen tracer diffusion
  along interfaces of strained {Y}$_2${O}$_3$/{YSZ} multilayers},}\ }\href
  {\doibase 10.1039/c2cp43231e} {\bibfield  {journal} {\bibinfo  {journal}
  {Phys. Chem. Chem. Phys.}\ }\textbf {\bibinfo {volume} {15}},\ \bibinfo
  {pages} {1944--1955} (\bibinfo {year} {2013})}\BibitemShut {NoStop}%
\bibitem [{\citenamefont {Shen}\ \emph {et~al.}(2014)\citenamefont {Shen},
  \citenamefont {Jiang},\ and\ \citenamefont {Hertz}}]{ShenEtAl_RSCAdv2014}%
  \BibitemOpen
  \bibfield  {author} {\bibinfo {author} {\bibfnamefont {Weida}\ \bibnamefont
  {Shen}}, \bibinfo {author} {\bibfnamefont {Jun}\ \bibnamefont {Jiang}}, \
  and\ \bibinfo {author} {\bibfnamefont {Joshua~L.}\ \bibnamefont {Hertz}},\
  }\bibfield  {title} {\enquote {\bibinfo {title} {Reduced ionic conductivity
  in biaxially compressed ceria},}\ }\href {\doibase 10.1039/C4RA00820K}
  {\bibfield  {journal} {\bibinfo  {journal} {RSC Adv.}\ }\textbf {\bibinfo
  {volume} {4}},\ \bibinfo {pages} {21625--21630} (\bibinfo {year}
  {2014})}\BibitemShut {NoStop}%
\bibitem [{\citenamefont {Fluri}\ \emph {et~al.}(2016)\citenamefont {Fluri},
  \citenamefont {Pergolesi}, \citenamefont {Roddatis}, \citenamefont {Wokaun},\
  and\ \citenamefont {Lippert}}]{FluriEtAl_NatureComm2016}%
  \BibitemOpen
  \bibfield  {author} {\bibinfo {author} {\bibfnamefont {Aline}\ \bibnamefont
  {Fluri}}, \bibinfo {author} {\bibfnamefont {Daniele}\ \bibnamefont
  {Pergolesi}}, \bibinfo {author} {\bibfnamefont {Vladimir}\ \bibnamefont
  {Roddatis}}, \bibinfo {author} {\bibfnamefont {Alexander}\ \bibnamefont
  {Wokaun}}, \ and\ \bibinfo {author} {\bibfnamefont {Thomas}\ \bibnamefont
  {Lippert}},\ }\bibfield  {title} {\enquote {\bibinfo {title} {In situ stress
  observation in oxide films and how tensile stress influences oxygen ion
  conduction},}\ }\href {\doibase 10.1038/ncomms10692} {\bibfield  {journal}
  {\bibinfo  {journal} {Nat. Comm.}\ }\textbf {\bibinfo {volume} {7}} (\bibinfo
  {year} {2016}),\ 10.1038/ncomms10692}\BibitemShut {NoStop}%
\bibitem [{\citenamefont {Ferrara}\ \emph {et~al.}(2016)\citenamefont
  {Ferrara}, \citenamefont {Eames}, \citenamefont {Islam},\ and\ \citenamefont
  {Tealdi}}]{FerraraEtAl_PhysChemChemPhys2016}%
  \BibitemOpen
  \bibfield  {author} {\bibinfo {author} {\bibfnamefont {Chiara}\ \bibnamefont
  {Ferrara}}, \bibinfo {author} {\bibfnamefont {Christopher}\ \bibnamefont
  {Eames}}, \bibinfo {author} {\bibfnamefont {M.~Saiful}\ \bibnamefont
  {Islam}}, \ and\ \bibinfo {author} {\bibfnamefont {Cristina}\ \bibnamefont
  {Tealdi}},\ }\bibfield  {title} {\enquote {\bibinfo {title} {Lattice strain
  effects on doping, hydration and proton transport in scheelite-type
  electrolytes for solid oxide fuel cells},}\ }\href@noop {} {\bibfield
  {journal} {\bibinfo  {journal} {Phys. Chem. Chem. Phys.}\ }\textbf {\bibinfo
  {volume} {18}},\ \bibinfo {pages} {29330--29336} (\bibinfo {year}
  {2016})}\BibitemShut {NoStop}%
\bibitem [{\citenamefont {Yan}\ \emph {et~al.}(2012)\citenamefont {Yan},
  \citenamefont {Wang}, \citenamefont {Xu},\ and\ \citenamefont
  {Ouyang}}]{YanEtAl_FunctMaterLett2012}%
  \BibitemOpen
  \bibfield  {author} {\bibinfo {author} {\bibfnamefont {Hui-Jun}\ \bibnamefont
  {Yan}}, \bibinfo {author} {\bibfnamefont {Zhi-Qiang}\ \bibnamefont {Wang}},
  \bibinfo {author} {\bibfnamefont {Bo}~\bibnamefont {Xu}}, \ and\ \bibinfo
  {author} {\bibfnamefont {Chuying}\ \bibnamefont {Ouyang}},\ }\bibfield
  {title} {\enquote {\bibinfo {title} {Strain induced enhanced migration of
  polaron and lithium ion in $\lambda$-{M}n{O}$_{2}$},}\ }\href@noop {}
  {\bibfield  {journal} {\bibinfo  {journal} {Funct. Mater. Lett.}\ }\textbf
  {\bibinfo {volume} {05}},\ \bibinfo {pages} {1250037--4} (\bibinfo {year}
  {2012})}\BibitemShut {NoStop}%
\bibitem [{\citenamefont {Lee}\ \emph {et~al.}(2012)\citenamefont {Lee},
  \citenamefont {Pennycook},\ and\ \citenamefont
  {Pantelides}}]{LeeEtAl_ApplPhysLett2012}%
  \BibitemOpen
  \bibfield  {author} {\bibinfo {author} {\bibfnamefont {Jaekwang}\
  \bibnamefont {Lee}}, \bibinfo {author} {\bibfnamefont {Stephen~J.}\
  \bibnamefont {Pennycook}}, \ and\ \bibinfo {author} {\bibfnamefont
  {Sokrates~T.}\ \bibnamefont {Pantelides}},\ }\bibfield  {title} {\enquote
  {\bibinfo {title} {Simultaneous enhancement of electronic and {L}i$^{+}$ ion
  conductivity in {L}i{F}e{P}{O}$_{4}$},}\ }\href@noop {} {\bibfield  {journal}
  {\bibinfo  {journal} {Appl. Phys. Lett.}\ }\textbf {\bibinfo {volume}
  {101}},\ \bibinfo {pages} {033901--5} (\bibinfo {year} {2012})}\BibitemShut
  {NoStop}%
\bibitem [{\citenamefont {Ning}\ \emph {et~al.}(2014)\citenamefont {Ning},
  \citenamefont {Li}, \citenamefont {Xu},\ and\ \citenamefont
  {Ouyang}}]{NingEtAl_SolStatIonics2014}%
  \BibitemOpen
  \bibfield  {author} {\bibinfo {author} {\bibfnamefont {Fanghua}\ \bibnamefont
  {Ning}}, \bibinfo {author} {\bibfnamefont {Shuai}\ \bibnamefont {Li}},
  \bibinfo {author} {\bibfnamefont {Bo}~\bibnamefont {Xu}}, \ and\ \bibinfo
  {author} {\bibfnamefont {Chuying}\ \bibnamefont {Ouyang}},\ }\bibfield
  {title} {\enquote {\bibinfo {title} {Strain tuned {L}i diffusion in
  {L}i{C}o{O}$_2$ material for {L}i ion batteries: A first principles study},}\
  }\href@noop {} {\bibfield  {journal} {\bibinfo  {journal} {Sol. Stat.
  Ionics}\ }\textbf {\bibinfo {volume} {263}},\ \bibinfo {pages} {46--48}
  (\bibinfo {year} {2014})}\BibitemShut {NoStop}%
\bibitem [{\citenamefont {Wei}\ \emph {et~al.}(2015)\citenamefont {Wei},
  \citenamefont {Ogawa}, \citenamefont {Fukumura}, \citenamefont {Hirose},\
  and\ \citenamefont {Hasegawa}}]{WeiEtAl_CrystGrowthDes2015}%
  \BibitemOpen
  \bibfield  {author} {\bibinfo {author} {\bibfnamefont {Jie}\ \bibnamefont
  {Wei}}, \bibinfo {author} {\bibfnamefont {Daisuke}\ \bibnamefont {Ogawa}},
  \bibinfo {author} {\bibfnamefont {Tomoteru}\ \bibnamefont {Fukumura}},
  \bibinfo {author} {\bibfnamefont {Yasushi}\ \bibnamefont {Hirose}}, \ and\
  \bibinfo {author} {\bibfnamefont {Tetsuya}\ \bibnamefont {Hasegawa}},\
  }\bibfield  {title} {\enquote {\bibinfo {title} {Epitaxial strain-controlled
  ionic conductivity in {L}i-ion solid electrolyte
  {L}i$_{0.33}${L}a$_{0.56}${T}i{O}$_3$ thin films},}\ }\href {\doibase
  10.1021/cg501834s} {\bibfield  {journal} {\bibinfo  {journal} {Crys. Growth
  \& Design}\ }\textbf {\bibinfo {volume} {15}},\ \bibinfo {pages} {2187--2191}
  (\bibinfo {year} {2015})}\BibitemShut {NoStop}%
\bibitem [{\citenamefont {Tealdi}\ \emph {et~al.}(2016)\citenamefont {Tealdi},
  \citenamefont {Heath},\ and\ \citenamefont {Islam}}]{Tealdi_JMaterChemA2016}%
  \BibitemOpen
  \bibfield  {author} {\bibinfo {author} {\bibfnamefont {Cristina}\
  \bibnamefont {Tealdi}}, \bibinfo {author} {\bibfnamefont {Jennifer}\
  \bibnamefont {Heath}}, \ and\ \bibinfo {author} {\bibfnamefont {M.~Saiful}\
  \bibnamefont {Islam}},\ }\bibfield  {title} {\enquote {\bibinfo {title}
  {Feeling the strain: enhancing ionic transport in olivine phosphate cathodes
  for {L}i- and {N}a-ion batteries through strain effects},}\ }\href {\doibase
  10.1039/C5TA09418F} {\bibfield  {journal} {\bibinfo  {journal} {J. Mater.
  Chem. A}\ }\textbf {\bibinfo {volume} {4}},\ \bibinfo {pages} {6998--7004}
  (\bibinfo {year} {2016})}\BibitemShut {NoStop}%
\bibitem [{\citenamefont {Moradabadi}\ and\ \citenamefont
  {Kaghazchi}(2017)}]{MoradabadiAndKaghazchi_PhysRevAppl2017}%
  \BibitemOpen
  \bibfield  {author} {\bibinfo {author} {\bibfnamefont {Ashkan}\ \bibnamefont
  {Moradabadi}}\ and\ \bibinfo {author} {\bibfnamefont {Payam}\ \bibnamefont
  {Kaghazchi}},\ }\bibfield  {title} {\enquote {\bibinfo {title} {Effect of
  strain on polaron hopping and electronic conductivity in bulk
  {L}i{C}o{O}$_2$},}\ }\href@noop {} {\bibfield  {journal} {\bibinfo  {journal}
  {Phys. Rev. Appl.}\ }\textbf {\bibinfo {volume} {7}},\ \bibinfo {pages}
  {064008--5} (\bibinfo {year} {2017})}\BibitemShut {NoStop}%
\bibitem [{\citenamefont {Jia}\ \emph {et~al.}(2017)\citenamefont {Jia},
  \citenamefont {Wang}, \citenamefont {Sun}, \citenamefont {Chen},
  \citenamefont {Guo},\ and\ \citenamefont {Gan}}]{JiaEtAl_RSCAdv2017}%
  \BibitemOpen
  \bibfield  {author} {\bibinfo {author} {\bibfnamefont {Mingzhen}\
  \bibnamefont {Jia}}, \bibinfo {author} {\bibfnamefont {Hongyan}\ \bibnamefont
  {Wang}}, \bibinfo {author} {\bibfnamefont {Zhandong}\ \bibnamefont {Sun}},
  \bibinfo {author} {\bibfnamefont {Yuanzheng}\ \bibnamefont {Chen}}, \bibinfo
  {author} {\bibfnamefont {Chunsheng}\ \bibnamefont {Guo}}, \ and\ \bibinfo
  {author} {\bibfnamefont {Liyong}\ \bibnamefont {Gan}},\ }\bibfield  {title}
  {\enquote {\bibinfo {title} {Exploring ion migration in
  {L}i$_2${M}n{S}i{O}$_4$ for {L}i-ion batteries through strain effects},}\
  }\href@noop {} {\bibfield  {journal} {\bibinfo  {journal} {RSC Adv.}\
  }\textbf {\bibinfo {volume} {7}},\ \bibinfo {pages} {26089--26096} (\bibinfo
  {year} {2017})}\BibitemShut {NoStop}%
\bibitem [{\citenamefont {Muralidharan}\ \emph {et~al.}(2017)\citenamefont
  {Muralidharan}, \citenamefont {Brock}, \citenamefont {Cohn}, \citenamefont
  {Schauben}, \citenamefont {Carter}, \citenamefont {Oakes}, \citenamefont
  {Walker},\ and\ \citenamefont {Pint}}]{MuralidharanEtAl_ACSNano2017}%
  \BibitemOpen
  \bibfield  {author} {\bibinfo {author} {\bibfnamefont {Nitin}\ \bibnamefont
  {Muralidharan}}, \bibinfo {author} {\bibfnamefont {Casey~N.}\ \bibnamefont
  {Brock}}, \bibinfo {author} {\bibfnamefont {Adam~P.}\ \bibnamefont {Cohn}},
  \bibinfo {author} {\bibfnamefont {Deanna}\ \bibnamefont {Schauben}}, \bibinfo
  {author} {\bibfnamefont {Rachel~E.}\ \bibnamefont {Carter}}, \bibinfo
  {author} {\bibfnamefont {Landon}\ \bibnamefont {Oakes}}, \bibinfo {author}
  {\bibfnamefont {D.~Greg}\ \bibnamefont {Walker}}, \ and\ \bibinfo {author}
  {\bibfnamefont {Cary~L.}\ \bibnamefont {Pint}},\ }\bibfield  {title}
  {\enquote {\bibinfo {title} {Tunable mechanochemistry of lithium battery
  electrodes},}\ }\href@noop {} {\bibfield  {journal} {\bibinfo  {journal} {ACS
  Nano}\ }\textbf {\bibinfo {volume} {11}},\ \bibinfo {pages} {6243--6251}
  (\bibinfo {year} {2017})}\BibitemShut {NoStop}%
\bibitem [{\citenamefont {Moradabadi}\ \emph {et~al.}(2017)\citenamefont
  {Moradabadi}, \citenamefont {Kaghazchi}, \citenamefont {Rohrer},\ and\
  \citenamefont {Albe}}]{MoradabadiEtAl_arXiv2017}%
  \BibitemOpen
  \bibfield  {author} {\bibinfo {author} {\bibfnamefont {Ashkan}\ \bibnamefont
  {Moradabadi}}, \bibinfo {author} {\bibfnamefont {Payam}\ \bibnamefont
  {Kaghazchi}}, \bibinfo {author} {\bibfnamefont {Jochen}\ \bibnamefont
  {Rohrer}}, \ and\ \bibinfo {author} {\bibfnamefont {Karsten}\ \bibnamefont
  {Albe}},\ }\bibfield  {title} {\enquote {\bibinfo {title} {Influence of
  elastic strain on the thermodynamics and kinetics of lithium vacancy in bulk
  {L}i{C}o{O}$_2$},}\ }\href@noop {} {\bibfield  {journal} {\bibinfo  {journal}
  {arXiv}\ } (\bibinfo {year} {2017})},\ \Eprint
  {http://arxiv.org/abs/1706.01709v1} {1706.01709v1} \BibitemShut {NoStop}%
\bibitem [{\citenamefont {Sagotra}\ and\ \citenamefont
  {Cazorla}(2017)}]{SagotraAndCazorla_ACSApplMaterInt2017}%
  \BibitemOpen
  \bibfield  {author} {\bibinfo {author} {\bibfnamefont {Arun~K.}\ \bibnamefont
  {Sagotra}}\ and\ \bibinfo {author} {\bibfnamefont {Claudio}\ \bibnamefont
  {Cazorla}},\ }\bibfield  {title} {\enquote {\bibinfo {title} {Stress-mediated
  enhancement of ionic conductivity in fast-ion conductors},}\ }\href@noop {}
  {\bibfield  {journal} {\bibinfo  {journal} {{ACS} Appl. Mater. Int.}\
  }\textbf {\bibinfo {volume} {9}},\ \bibinfo {pages} {38773--38783} (\bibinfo
  {year} {2017})}\BibitemShut {NoStop}%
\bibitem [{\citenamefont {Mees}\ \emph {et~al.}(2014)\citenamefont {Mees},
  \citenamefont {Pourtois}, \citenamefont {Rosciano}, \citenamefont {Put},
  \citenamefont {Vereecken},\ and\ \citenamefont
  {Stesmans}}]{MeesEtAl_PhysChemChemPhys2014}%
  \BibitemOpen
  \bibfield  {author} {\bibinfo {author} {\bibfnamefont {Maarten~J.}\
  \bibnamefont {Mees}}, \bibinfo {author} {\bibfnamefont {Geoffrey}\
  \bibnamefont {Pourtois}}, \bibinfo {author} {\bibfnamefont {Fabio}\
  \bibnamefont {Rosciano}}, \bibinfo {author} {\bibfnamefont {Brecht}\
  \bibnamefont {Put}}, \bibinfo {author} {\bibfnamefont {Philippe~M.}\
  \bibnamefont {Vereecken}}, \ and\ \bibinfo {author} {\bibfnamefont {Andr\'e}\
  \bibnamefont {Stesmans}},\ }\bibfield  {title} {\enquote {\bibinfo {title}
  {First-principles material modeling of solid-state electrolytes with the
  spinel structure},}\ }\href@noop {} {\bibfield  {journal} {\bibinfo
  {journal} {Phys. Chem. Chem. Phys.}\ }\textbf {\bibinfo {volume} {16}},\
  \bibinfo {pages} {5399--5406} (\bibinfo {year} {2014})}\BibitemShut {NoStop}%
\bibitem [{\citenamefont {Henkelman}\ \emph {et~al.}(2000)\citenamefont
  {Henkelman}, \citenamefont {Uberuaga},\ and\ \citenamefont
  {J\'{o}nsson}}]{HenkelmanEtAl_JChemPhys2000}%
  \BibitemOpen
  \bibfield  {author} {\bibinfo {author} {\bibfnamefont {Graeme}\ \bibnamefont
  {Henkelman}}, \bibinfo {author} {\bibfnamefont {Blas~P.}\ \bibnamefont
  {Uberuaga}}, \ and\ \bibinfo {author} {\bibfnamefont {Hannes}\ \bibnamefont
  {J\'{o}nsson}},\ }\bibfield  {title} {\enquote {\bibinfo {title} {A climbing
  image nudged elastic band method for finding saddle points and minimum energy
  paths},}\ }\href@noop {} {\bibfield  {journal} {\bibinfo  {journal} {J. Chem.
  Phys.}\ }\textbf {\bibinfo {volume} {113}},\ \bibinfo {pages} {9901--9904}
  (\bibinfo {year} {2000})}\BibitemShut {NoStop}%
\bibitem [{\citenamefont {Allimi}\ \emph {et~al.}(2008)\citenamefont {Allimi},
  \citenamefont {Aindow},\ and\ \citenamefont
  {Alpay}}]{AllimiEtAl_ApplPhysLett2008}%
  \BibitemOpen
  \bibfield  {author} {\bibinfo {author} {\bibfnamefont {B.~S.}\ \bibnamefont
  {Allimi}}, \bibinfo {author} {\bibfnamefont {M.}~\bibnamefont {Aindow}}, \
  and\ \bibinfo {author} {\bibfnamefont {S.~P.}\ \bibnamefont {Alpay}},\
  }\bibfield  {title} {\enquote {\bibinfo {title} {Thickness dependence of
  electronic phase transitions in epitaxial {V}$_2${O}$_3$ films on $(0001)$
  {L}i{T}a{O}$_3$},}\ }\href@noop {} {\bibfield  {journal} {\bibinfo  {journal}
  {Appl. Phys. Lett.}\ }\textbf {\bibinfo {volume} {93}},\ \bibinfo {pages}
  {112109} (\bibinfo {year} {2008})}\BibitemShut {NoStop}%
\bibitem [{\citenamefont {Kresse}\ and\ \citenamefont
  {Furthm{\"{u}}ller}(1996)}]{KresseAndFurthmuller_PhysRevB1996}%
  \BibitemOpen
  \bibfield  {author} {\bibinfo {author} {\bibfnamefont {G.}~\bibnamefont
  {Kresse}}\ and\ \bibinfo {author} {\bibfnamefont {J.}~\bibnamefont
  {Furthm{\"{u}}ller}},\ }\bibfield  {title} {\enquote {\bibinfo {title}
  {Efficient iterative schemes for ab initio total-energy calculations using a
  plane-wave basis set},}\ }\href {\doibase 10.1103/PhysRevB.54.11169}
  {\bibfield  {journal} {\bibinfo  {journal} {Phys. Rev. B}\ }\textbf {\bibinfo
  {volume} {54}},\ \bibinfo {pages} {11169--11186} (\bibinfo {year}
  {1996})}\BibitemShut {NoStop}%
\bibitem [{\citenamefont {Kresse}\ and\ \citenamefont
  {Furthm{\"u}ller}(1996)}]{KresseAndFurthmuller_CompMaterSci1996}%
  \BibitemOpen
  \bibfield  {author} {\bibinfo {author} {\bibfnamefont {G.}~\bibnamefont
  {Kresse}}\ and\ \bibinfo {author} {\bibfnamefont {J.}~\bibnamefont
  {Furthm{\"u}ller}},\ }\bibfield  {title} {\enquote {\bibinfo {title}
  {Efficiency of ab-initio total energy calculations for metals and
  semiconductors using a plane-wave basis set},}\ }\href@noop {} {\bibfield
  {journal} {\bibinfo  {journal} {Comp. Mater. Sci.}\ }\textbf {\bibinfo
  {volume} {6}},\ \bibinfo {pages} {15--50} (\bibinfo {year}
  {1996})}\BibitemShut {NoStop}%
\bibitem [{\citenamefont {Perdew}\ \emph {et~al.}(2008)\citenamefont {Perdew},
  \citenamefont {Ruzsinszky}, \citenamefont {Csonka}, \citenamefont {Vydrov},
  \citenamefont {Scuseria}, \citenamefont {Constantin}, \citenamefont {Zhou},\
  and\ \citenamefont {Burke}}]{PerdewEtAl_PhysRevLett2008}%
  \BibitemOpen
  \bibfield  {author} {\bibinfo {author} {\bibfnamefont {John~P.}\ \bibnamefont
  {Perdew}}, \bibinfo {author} {\bibfnamefont {Adrienn}\ \bibnamefont
  {Ruzsinszky}}, \bibinfo {author} {\bibfnamefont {G\'abor~I.}\ \bibnamefont
  {Csonka}}, \bibinfo {author} {\bibfnamefont {Oleg~A.}\ \bibnamefont
  {Vydrov}}, \bibinfo {author} {\bibfnamefont {Gustavo~E.}\ \bibnamefont
  {Scuseria}}, \bibinfo {author} {\bibfnamefont {Lucian~A.}\ \bibnamefont
  {Constantin}}, \bibinfo {author} {\bibfnamefont {Xiaolan}\ \bibnamefont
  {Zhou}}, \ and\ \bibinfo {author} {\bibfnamefont {Kieron}\ \bibnamefont
  {Burke}},\ }\bibfield  {title} {\enquote {\bibinfo {title} {Restoring the
  density-gradient expansion for exchange in solids and surfaces},}\ }\href
  {\doibase 10.1103/PhysRevLett.100.136406} {\bibfield  {journal} {\bibinfo
  {journal} {Phys. Rev. Lett.}\ }\textbf {\bibinfo {volume} {100}},\ \bibinfo
  {pages} {136406} (\bibinfo {year} {2008})}\BibitemShut {NoStop}%
\bibitem [{\citenamefont {Bl{\"{o}}chl}(1994)}]{Blochl_PhysRevB1994}%
  \BibitemOpen
  \bibfield  {author} {\bibinfo {author} {\bibfnamefont {P.~E.}\ \bibnamefont
  {Bl{\"{o}}chl}},\ }\bibfield  {title} {\enquote {\bibinfo {title} {Projector
  augmented-wave method},}\ }\href@noop {} {\bibfield  {journal} {\bibinfo
  {journal} {Phys. Rev. B}\ }\textbf {\bibinfo {volume} {50}},\ \bibinfo
  {pages} {17953--17979} (\bibinfo {year} {1994})}\BibitemShut {NoStop}%
\bibitem [{\citenamefont {Sickafus}\ \emph {et~al.}(1999)\citenamefont
  {Sickafus}, \citenamefont {Wills},\ and\ \citenamefont
  {Grimes}}]{SickafusEtAl_JAmCermSoc1999}%
  \BibitemOpen
  \bibfield  {author} {\bibinfo {author} {\bibfnamefont {Kurt~E.}\ \bibnamefont
  {Sickafus}}, \bibinfo {author} {\bibfnamefont {John~M.}\ \bibnamefont
  {Wills}}, \ and\ \bibinfo {author} {\bibfnamefont {Norman~W.}\ \bibnamefont
  {Grimes}},\ }\bibfield  {title} {\enquote {\bibinfo {title} {Structure of
  spinel},}\ }\href {\doibase 10.1111/j.1151-2916.1999.tb02241.x} {\bibfield
  {journal} {\bibinfo  {journal} {J. Am. Ceram. Soc.}\ }\textbf {\bibinfo
  {volume} {82}},\ \bibinfo {pages} {3279--3292} (\bibinfo {year}
  {1999})}\BibitemShut {NoStop}%
\bibitem [{\citenamefont {Hill}\ \emph {et~al.}(1979)\citenamefont {Hill},
  \citenamefont {Craig},\ and\ \citenamefont
  {Gibbs}}]{HillEtAl_PhysChemMiner1979}%
  \BibitemOpen
  \bibfield  {author} {\bibinfo {author} {\bibfnamefont {Roderick~J.}\
  \bibnamefont {Hill}}, \bibinfo {author} {\bibfnamefont {James~R.}\
  \bibnamefont {Craig}}, \ and\ \bibinfo {author} {\bibfnamefont {G.~V.}\
  \bibnamefont {Gibbs}},\ }\bibfield  {title} {\enquote {\bibinfo {title}
  {Systematics of the spinel structure type},}\ }\href {\doibase
  10.1007/BF00307535} {\bibfield  {journal} {\bibinfo  {journal} {Phys. Chem.
  Mater.}\ }\textbf {\bibinfo {volume} {4}},\ \bibinfo {pages} {317--339}
  (\bibinfo {year} {1979})}\BibitemShut {NoStop}%
\bibitem [{\citenamefont {Morgan}(2017)}]{Morgan_JOpenSourceSoftware_bsym}%
  \BibitemOpen
  \bibfield  {author} {\bibinfo {author} {\bibfnamefont {Benjamin~J.}\
  \bibnamefont {Morgan}},\ }\bibfield  {title} {\enquote {\bibinfo {title}
  {bsym: A basic symmetry module},}\ }\href {\doibase 10.21105/joss.00370}
  {\bibfield  {journal} {\bibinfo  {journal} {The Journal of Open Source
  Software}\ }\textbf {\bibinfo {volume} {2}} (\bibinfo {year} {2017}),\
  10.21105/joss.00370}\BibitemShut {NoStop}%
\bibitem [{\citenamefont {Ronci}\ \emph {et~al.}(2002)\citenamefont {Ronci},
  \citenamefont {Reale}, \citenamefont {Scrosati}, \citenamefont {Panero},
  \citenamefont {Rossi~Albertini}, \citenamefont {Perfetti}, \citenamefont
  {di~Michiel},\ and\ \citenamefont {Merino}}]{RonciEtAl_JPhysChemB2002}%
  \BibitemOpen
  \bibfield  {author} {\bibinfo {author} {\bibfnamefont {F.}~\bibnamefont
  {Ronci}}, \bibinfo {author} {\bibfnamefont {P.}~\bibnamefont {Reale}},
  \bibinfo {author} {\bibfnamefont {B.}~\bibnamefont {Scrosati}}, \bibinfo
  {author} {\bibfnamefont {S.}~\bibnamefont {Panero}}, \bibinfo {author}
  {\bibfnamefont {V.}~\bibnamefont {Rossi~Albertini}}, \bibinfo {author}
  {\bibfnamefont {P.}~\bibnamefont {Perfetti}}, \bibinfo {author}
  {\bibfnamefont {M.}~\bibnamefont {di~Michiel}}, \ and\ \bibinfo {author}
  {\bibfnamefont {J.~M.}\ \bibnamefont {Merino}},\ }\bibfield  {title}
  {\enquote {\bibinfo {title} {High-resolution in-situ structural measurements
  of the {L}i$_{4/3}${T}i$_{5/3}${O}$_{4
  }${\textquotedblleft}zero-strain{\textquotedblright} insertion material},}\
  }\href@noop {} {\bibfield  {journal} {\bibinfo  {journal} {J. Phys. Chem. B}\
  }\textbf {\bibinfo {volume} {106}},\ \bibinfo {pages} {3082--3086} (\bibinfo
  {year} {2002})}\BibitemShut {NoStop}%
\bibitem [{\citenamefont {Morgan}\ \emph {et~al.}(2016)\citenamefont {Morgan},
  \citenamefont {Carrasco},\ and\ \citenamefont
  {Teobaldi}}]{MorganEtAl_JMaterChemA2016}%
  \BibitemOpen
  \bibfield  {author} {\bibinfo {author} {\bibfnamefont {Benjamin~J.}\
  \bibnamefont {Morgan}}, \bibinfo {author} {\bibfnamefont {Javier}\
  \bibnamefont {Carrasco}}, \ and\ \bibinfo {author} {\bibfnamefont {Gilberto}\
  \bibnamefont {Teobaldi}},\ }\bibfield  {title} {\enquote {\bibinfo {title}
  {Variation in surface energy and reduction drive of a metal oxide lithium-ion
  anode with stoichiometry: a {DFT} study of lithium titanate spinel
  surfaces},}\ }\href {\doibase 10.1039/c6ta05980e} {\bibfield  {journal}
  {\bibinfo  {journal} {J. Mater. Chem. A}\ }\textbf {\bibinfo {volume} {4}},\
  \bibinfo {pages} {17180--17192} (\bibinfo {year} {2016})}\BibitemShut
  {NoStop}%
\bibitem [{\citenamefont {Morgan}\ and\ \citenamefont
  {Watson}(2010)}]{MorganAndWatson_PhysRevB2010}%
  \BibitemOpen
  \bibfield  {author} {\bibinfo {author} {\bibfnamefont {Benjamin~J.}\
  \bibnamefont {Morgan}}\ and\ \bibinfo {author} {\bibfnamefont {Graeme~W.}\
  \bibnamefont {Watson}},\ }\bibfield  {title} {\enquote {\bibinfo {title}
  {{GGA}$+{U}$ description of lithium intercalation into anatase
  {T}i{O}$_2$},}\ }\href {\doibase 10.1103/PhysRevB.82.144119} {\bibfield
  {journal} {\bibinfo  {journal} {Phys. Rev. B}\ }\textbf {\bibinfo {volume}
  {82}},\ \bibinfo {pages} {144119} (\bibinfo {year} {2010})}\BibitemShut
  {NoStop}%
\bibitem [{\citenamefont {Tompsett}\ \emph {et~al.}(2013)\citenamefont
  {Tompsett}, \citenamefont {Parker}, \citenamefont {Bruce},\ and\
  \citenamefont {Islam}}]{TompsettEtAl_ChemMater2013}%
  \BibitemOpen
  \bibfield  {author} {\bibinfo {author} {\bibfnamefont {David~A.}\
  \bibnamefont {Tompsett}}, \bibinfo {author} {\bibfnamefont {Steve~C.}\
  \bibnamefont {Parker}}, \bibinfo {author} {\bibfnamefont {Peter~G.}\
  \bibnamefont {Bruce}}, \ and\ \bibinfo {author} {\bibfnamefont {M.~Saiful}\
  \bibnamefont {Islam}},\ }\bibfield  {title} {\enquote {\bibinfo {title}
  {Nanostructuring of $\beta$-{M}n{O}$_2$: The important role of surface to
  bulk ion migration},}\ }\href {\doibase 10.1021/cm303295f} {\bibfield
  {journal} {\bibinfo  {journal} {Chem. Mater.}\ }\textbf {\bibinfo {volume}
  {25}},\ \bibinfo {pages} {536--541} (\bibinfo {year} {2013})}\BibitemShut
  {NoStop}%
\bibitem [{\citenamefont {Rong}\ \emph {et~al.}(2016)\citenamefont {Rong},
  \citenamefont {Kitchaev}, \citenamefont {Canepa}, \citenamefont {Huang},\
  and\ \citenamefont {Ceder}}]{RongEtAl_JChemPhys2016}%
  \BibitemOpen
  \bibfield  {author} {\bibinfo {author} {\bibfnamefont {Ziqin}\ \bibnamefont
  {Rong}}, \bibinfo {author} {\bibfnamefont {Daniil}\ \bibnamefont {Kitchaev}},
  \bibinfo {author} {\bibfnamefont {Pieremanuele}\ \bibnamefont {Canepa}},
  \bibinfo {author} {\bibfnamefont {Wenxuan}\ \bibnamefont {Huang}}, \ and\
  \bibinfo {author} {\bibfnamefont {Gerbrand}\ \bibnamefont {Ceder}},\
  }\bibfield  {title} {\enquote {\bibinfo {title} {An efficient algorithm for
  finding the minimum energy path for cation migration in ionic materials},}\
  }\href@noop {} {\bibfield  {journal} {\bibinfo  {journal} {J. Chem. Phys.}\
  }\textbf {\bibinfo {volume} {145}},\ \bibinfo {pages} {074112--9} (\bibinfo
  {year} {2016})}\BibitemShut {NoStop}%
\bibitem [{\citenamefont {O'Rourke}\ and\ \citenamefont
  {Morgan}(2017{\natexlab{a}})}]{ORourkeAndMorgan_SpinelNEBDataset2017}%
  \BibitemOpen
  \bibfield  {author} {\bibinfo {author} {\bibfnamefont {Conn}\ \bibnamefont
  {O'Rourke}}\ and\ \bibinfo {author} {\bibfnamefont {Benjamin~J.}\
  \bibnamefont {Morgan}},\ }\href {\doibase 10.15125/BATH-00438} {\enquote
  {\bibinfo {title} {{DFT} dataset for ``{I}nterfacial strain effects on
  lithium diffusion pathways in the spinel solid electrolyte {L}i-doped
  {M}g{A}l$_2${O}$_4$''},}\ }\bibinfo {howpublished}
  {https://doi.org/10.15125/BATH-00438} (\bibinfo {year}
  {2017}{\natexlab{a}})\BibitemShut {NoStop}%
\bibitem [{\citenamefont {O'Rourke}\ and\ \citenamefont
  {Morgan}(2017{\natexlab{b}})}]{ORourkeAndMorgan_SpinelAnalysisNotebook2017}%
  \BibitemOpen
  \bibfield  {author} {\bibinfo {author} {\bibfnamefont {Conn}\ \bibnamefont
  {O'Rourke}}\ and\ \bibinfo {author} {\bibfnamefont {Benjamin~J.}\
  \bibnamefont {Morgan}},\ }\href {\doibase 10.5281/zenodo.1069417} {\enquote
  {\bibinfo {title} {Data analysis for ``{I}nterfacial strain effects on
  lithium diffusion pathways in the spinel solid electrolyte {L}i-doped
  {M}g{A}l$_2${O}$_4$''},}\ }\bibinfo {howpublished}
  {https://github.com/bjmorgan/data\_NEB\_spinel} (\bibinfo {year}
  {2017}{\natexlab{b}})\BibitemShut {NoStop}%
\bibitem [{\citenamefont {Hunter}(2007)}]{Hunter_matplotlib2007}%
  \BibitemOpen
  \bibfield  {author} {\bibinfo {author} {\bibfnamefont {J.~D.}\ \bibnamefont
  {Hunter}},\ }\bibfield  {title} {\enquote {\bibinfo {title} {Matplotlib: A
  {2D} graphics environment},}\ }\href {\doibase 10.1109/MCSE.2007.55}
  {\bibfield  {journal} {\bibinfo  {journal} {Computing In Science \&
  Engineering}\ }\textbf {\bibinfo {volume} {9}},\ \bibinfo {pages} {90--95}
  (\bibinfo {year} {2007})}\BibitemShut {NoStop}%
\bibitem [{\citenamefont {Kluyver}\ \emph {et~al.}(2016)\citenamefont
  {Kluyver}, \citenamefont {Ragan-Kelley}, \citenamefont {P{\'e}rez},
  \citenamefont {Granger}, \citenamefont {Bussonnier}, \citenamefont
  {Frederic}, \citenamefont {Kelley}, \citenamefont {Hamrick}, \citenamefont
  {Grout}, \citenamefont {Corlay}, \citenamefont {Ivanov}, \citenamefont
  {Avila}, \citenamefont {Abdalla},\ and\ \citenamefont
  {Willing}}]{KluyverEtAl_Jupyter2016}%
  \BibitemOpen
  \bibfield  {author} {\bibinfo {author} {\bibfnamefont {Thomas}\ \bibnamefont
  {Kluyver}}, \bibinfo {author} {\bibfnamefont {Benjamin}\ \bibnamefont
  {Ragan-Kelley}}, \bibinfo {author} {\bibfnamefont {Fernando}\ \bibnamefont
  {P{\'e}rez}}, \bibinfo {author} {\bibfnamefont {Brian}\ \bibnamefont
  {Granger}}, \bibinfo {author} {\bibfnamefont {Matthias}\ \bibnamefont
  {Bussonnier}}, \bibinfo {author} {\bibfnamefont {Jonathan}\ \bibnamefont
  {Frederic}}, \bibinfo {author} {\bibfnamefont {Kyle}\ \bibnamefont {Kelley}},
  \bibinfo {author} {\bibfnamefont {Jessica}\ \bibnamefont {Hamrick}}, \bibinfo
  {author} {\bibfnamefont {Jason}\ \bibnamefont {Grout}}, \bibinfo {author}
  {\bibfnamefont {Sylvain}\ \bibnamefont {Corlay}}, \bibinfo {author}
  {\bibfnamefont {Paul}\ \bibnamefont {Ivanov}}, \bibinfo {author}
  {\bibfnamefont {Dami{\'a}n}\ \bibnamefont {Avila}}, \bibinfo {author}
  {\bibfnamefont {Safia}\ \bibnamefont {Abdalla}}, \ and\ \bibinfo {author}
  {\bibfnamefont {Carol}\ \bibnamefont {Willing}},\ }\bibfield  {title}
  {\enquote {\bibinfo {title} {Jupyter notebooks -- a publishing format for
  reproducible computational workflows},}\ }in\ \href@noop {} {\emph {\bibinfo
  {booktitle} {Positioning and Power in Academic Publishing: Players, Agents
  and Agendas}}},\ \bibinfo {editor} {edited by\ \bibinfo {editor}
  {\bibfnamefont {F.}~\bibnamefont {Loizides}}\ and\ \bibinfo {editor}
  {\bibfnamefont {B.}~\bibnamefont {Schmidt}}}\ (\bibinfo {organization} {IOS
  Press},\ \bibinfo {year} {2016})\ pp.\ \bibinfo {pages} {87 --
  90}\BibitemShut {NoStop}%
\bibitem [{\citenamefont {Momma}\ and\ \citenamefont
  {Izumi}(2011)}]{MommaEtAlJApplCryst2011}%
  \BibitemOpen
  \bibfield  {author} {\bibinfo {author} {\bibfnamefont {Koichi}\ \bibnamefont
  {Momma}}\ and\ \bibinfo {author} {\bibfnamefont {Fujio}\ \bibnamefont
  {Izumi}},\ }\bibfield  {title} {\enquote {\bibinfo {title} {\texttt{VESTA3}
  for three-dimensional visualization of crystal, volumetric and morphology
  data},}\ }\href {\doibase 10.1107/S0021889811038970} {\bibfield  {journal}
  {\bibinfo  {journal} {J. Appl. Cryst.}\ }\textbf {\bibinfo {volume} {44}},\
  \bibinfo {pages} {1272--1276} (\bibinfo {year} {2011})}\BibitemShut {NoStop}%
\bibitem [{\citenamefont {Put}\ \emph {et~al.}(2015)\citenamefont {Put},
  \citenamefont {Vereecken}, \citenamefont {Mees}, \citenamefont {Rosciano},
  \citenamefont {Radu},\ and\ \citenamefont
  {Stesmans}}]{PutEtAl_PhysChemChemPhys2015}%
  \BibitemOpen
  \bibfield  {author} {\bibinfo {author} {\bibfnamefont {Brecht}\ \bibnamefont
  {Put}}, \bibinfo {author} {\bibfnamefont {Philippe~M.}\ \bibnamefont
  {Vereecken}}, \bibinfo {author} {\bibfnamefont {Maarten~J.}\ \bibnamefont
  {Mees}}, \bibinfo {author} {\bibfnamefont {Fabio}\ \bibnamefont {Rosciano}},
  \bibinfo {author} {\bibfnamefont {Iuliana~P.}\ \bibnamefont {Radu}}, \ and\
  \bibinfo {author} {\bibfnamefont {Andre}\ \bibnamefont {Stesmans}},\
  }\bibfield  {title} {\enquote {\bibinfo {title} {Characterization of thin
  films of the solid electrolyte {L}i$_x${M}g$_{1-2x}${A}l$_{2+x}$o$_4$ ($x =
  0, 0.05, 0.15, 0.25$)},}\ }\href@noop {} {\bibfield  {journal} {\bibinfo
  {journal} {Phys. Chem. Chem. Phys.}\ }\textbf {\bibinfo {volume} {17}},\
  \bibinfo {pages} {29045--29056} (\bibinfo {year} {2015})}\BibitemShut
  {NoStop}%
\bibitem [{\citenamefont {Morgan}\ and\ \citenamefont
  {Madden}(2014)}]{MorganAndMadden_PhysRevB2014}%
  \BibitemOpen
  \bibfield  {author} {\bibinfo {author} {\bibfnamefont {Benjamin~J.}\
  \bibnamefont {Morgan}}\ and\ \bibinfo {author} {\bibfnamefont {Paul~A.}\
  \bibnamefont {Madden}},\ }\bibfield  {title} {\enquote {\bibinfo {title}
  {Frenkel polarisation of coherent interfaces in fluorite heterostructures},}\
  }\href {\doibase 10.1103/PhysRevB.89.054304} {\bibfield  {journal} {\bibinfo
  {journal} {Phys. Rev. B}\ }\textbf {\bibinfo {volume} {89}} (\bibinfo {year}
  {2014}),\ 10.1103/PhysRevB.89.054304}\BibitemShut {NoStop}%
\bibitem [{\citenamefont {Leung}\ and\ \citenamefont
  {Leenheer}(2015)}]{LeungAndLeenheer_JPhysChemC2015}%
  \BibitemOpen
  \bibfield  {author} {\bibinfo {author} {\bibfnamefont {Kevin}\ \bibnamefont
  {Leung}}\ and\ \bibinfo {author} {\bibfnamefont {Andrew}\ \bibnamefont
  {Leenheer}},\ }\bibfield  {title} {\enquote {\bibinfo {title} {How voltage
  drops are manifested by lithium ion configurations at interfaces and in thin
  films on battery electrodes},}\ }\href@noop {} {\bibfield  {journal}
  {\bibinfo  {journal} {J. Phys. Chem. C}\ }\textbf {\bibinfo {volume} {119}},\
  \bibinfo {pages} {10234--10246} (\bibinfo {year} {2015})}\BibitemShut
  {NoStop}%
\end{thebibliography}%

\end{document}